\documentclass{IEEEtran}
\usepackage{amsmath,amssymb,amsfonts}
\usepackage{graphicx}
\usepackage{amsfonts,helvet}
\usepackage{fancyhdr}
\usepackage{textcomp}
\usepackage{xcolor}
\usepackage{caption}
\usepackage{subcaption}
\usepackage{tabularx}
\usepackage[english]{babel}
\usepackage{amsthm}
\usepackage{tabularx}
\usepackage{multirow}
\usepackage[noadjust]{cite}
\usepackage{mcite}

\usepackage[linesnumbered,ruled]{algorithm2e}
\SetKwInput{KwInput}{Initialize}
\SetKwInput{KwP}{Power measurement \& Decoding order}
\SetKwInput{KwOutput}{Output}
\SetKwProg{Fn}{Stage 1}{}{}
\SetKwProg{Fn}{Stage 2}{}{}
\newtheorem{remark}{Remark}

\begin{document}
\title{AoA-based Position and Orientation Estimation Using Lens MIMO in Cooperative Vehicle-to-Vehicle Systems}
\author{Joo-Hyun Jo,~\IEEEmembership{Student Member,~IEEE,}
        Jae-Nam Shim,~\IEEEmembership{Member,~IEEE,}
        Byoungnam (Klaus) Kim,~\IEEEmembership{Member,~IEEE,}
        Chan-Byoung Chae,~\IEEEmembership{Fellow,~IEEE,}
        and~Dong Ku Kim,~\IEEEmembership{Senior Member,~IEEE}
\thanks{J.-H. Jo, C.-B. Chae and D. K. Kim are with Yonsei University, Seoul, Korea (e-mail: \{joohyun\_jo, cbchae, dkkim\}@yonsei.ac.kr). J.-N. Shim is with LG Electronics, Seoul, Korea (e-mail: jaenam.shim@lge.com). B. Kim is with SensorView, Ltd., Gyeonggi-do, 13493, South Korea (e-mail: klaus.kim@sensor-view.com).

This work was supported by the National Research Foundation of Korea (NRF) Grant through the Ministry of Science and ICT (MSIT), under Grant 2022R1A5A1027646, and the Ministry of Oceans and Fisheries, under Grant 20210636.

Source codes are available at https://www.cbchae.org/}}
\maketitle
\begin{abstract}

Positioning accuracy is a critical requirement for vehicle-to-everything (V2X) use cases. Therefore, this paper derives the theoretical limits of estimation for the position and orientation of vehicles in a cooperative vehicle-to-vehicle (V2V) scenario, using a lens-based multiple-input multiple-output (lens-MIMO) system. Following this, we analyze the Cram$\acute{\text{e}}$r-Rao lower bounds (CRLBs) of the position and orientation estimation and explore a received signal model of a lens-MIMO for the particular angle of arrival (AoA) estimation with a V2V geometric model. Further, we propose a lower complexity AoA estimation technique exploiting the unique characteristics of the lens-MIMO for a single target vehicle; as a result, its estimation scheme is effectively extended by the successive interference cancellation (SIC) method for multiple target vehicles. Given these AoAs, we investigate the lens-MIMO estimation capability for the positions and orientations of vehicles. Subsequently, we prove that the lens-MIMO outperforms a conventional uniform linear array (ULA) in a certain configuration of a lens's structure. Finally, we confirm that the proposed localization algorithm is superior to ULA's CRLB as the resolution of the lens increases in spite of the lower complexity.
\end{abstract}

\begin{IEEEkeywords}
Position and orientation, Cram$\acute{\text{e}}$r-Rao lower bound (CRLB), lens-based multiple-input multiple-output (lens-MIMO), cooperative localization, AoA estimation.
\end{IEEEkeywords}
\section{Introduction}\label{sec1}
\IEEEPARstart{H}{igher} accuracy positioning is one of the essential requirements for various 5G vehicle-to-everything (V2X) advanced driving use cases\textemdash For instance, in location-aware communications at the street intersection depicted in the 3rd generation partnership project (3GPP) Rel-16 \cite{b11,b1,b13}. Interestingly, the 5G automotive association (5GAA) highlighted highly accurate localization requirements as one of the key indicators for the autonomous vertical industries \cite{b5}. More so, the acquisition of the orientation information of surrounding vehicles helps to predict a vehicle’s maneuvering efficiently to make autonomous driving vehicles safer in terms of the planned movements of all of the surrounding vehicles \cite{b2}. As a result, cooperative localization based on the millimeter-wave (mmWave) multiple-input-multiple-output (MIMO) in the vehicle-to-vehicle (V2V) systems have been investigated to provide better estimates of the position and orientation of vehicles \cite{b3}. Specifically, the mmWave can propagate along the line-of-sight (LoS) path with little reflection and scattering, which is beneficial to higher-precision positioning. However, this requires a significant computational burden and increased energy consumption in the case of a large number of antennas and higher radio frequencies (RFs), which are widely implemented with a hybrid arrangement of analog and digital precoders. Hence, to overcome these inherent challenges, the lens-based multiple-input multiple-output (lens-MIMO) has been explored in vehicular applications \cite{b4, b12, add2}.

Meanwhile, several positioning techniques within the radio access network (RAN) are standardized to meet the requirements of V2X services. Specifically, the infrastructure plays a central role in determining vehicular locations in terms of the measurement of angles and distances from other vehicles \cite{b17}. However, due to the limit of acquiring the real-estate for base-stations (BS) or road side units (RSUs), and more importantly, the potential path-blockage between the gNodeBs and vehicles, the cooperative localization among the vehicles has received extensive interest in V2X positioning use cases \cite{loc1,loc2,loc3}. Furthermore, the cooperative positioning schemes using direct channels among vehicles, such as the sidelink, which is currently being examined in the 3GPP Rel-18 work item \cite{b18}, will help meet the requirements of localization accuracy. Unlike the conventional uniform linear array (ULA), whose error bounds and localization performance have been widely investigated \cite{b14, b15}, the lens-MIMO still requires further examination to determine the feasibility of its localization capability, especially with lower complexity in V2V scenarios. 

Furthermore, the accurate estimation of channel parameters between vehicles is crucial to enhancing the overall performance of the position and orientation of vehicles. In lens-MIMO, the lens is inherently capable of focusing the energy of all of the wavefronts, which are incident to the lens surface with a single angle of arrival (AoA), into one focal point. Therefore, each focal point represents a specific angle of arrival, and the lens-MIMO would spatially sample the received signal according to its antenna placement. This capability allows for the physical separation of received signals corresponding to different incident angles across the antenna elements. This facilitates a certain level of inter-path interference suppression among these signals, where the extent of suppression gain depends on the disparities in their AoAs. Consequently, this feature contributes to the enhancement of multiple AoA estimation. It should be noted that the placement of the antenna elements on the focal region is associated with the unique arriving directions to effectively receive the maximum received signal power, which facilitates a simpler AoA estimation implementation \cite{b16}. In \cite{est6}, the authors investigated the maximum likelihood (ML) technique, exploiting the received signal of the lens-MIMO to estimate the AoA. Additionally, to reduce the computational complexity of the AoA estimation, the authors in \cite{est4, est1}, and \cite{est2} have exploited a sparse structure of the mmWave channel and the energy-focusing property of the lens-MIMO. Precisely, the key idea of these schemes is to efficiently utilize the sparsity of the mmWave MIMO channel by using only some antenna elements of the strongly received energy. However, the aforementioned works still constitute a high computational burden either in calculating the covariance matrix of the reduced spatial channel or in searching for the AoA with a large-sized dictionary of the precomputed received array response vectors. Therefore, to the best of our knowledge, the lens-MIMO AoA estimation technique still requires further complexity reduction to be useful for practical applications. This concern has motivated us to further explore the lens-MIMO localization capability in terms of not only a simpler AoA estimation implementation scheme but also its performance superiority over ULA.

In this paper, we propose a lower complexity AoA
estimation algorithm that exploits the inherent structure of the lens-MIMO. Specifically, its estimate will be utilized in the analysis of the Cram$\acute{\text{e}}$r-Rao lower bounds (CRLBs) for the position and orientation estimation. Additionally, we investigate the cooperative
V2V localization performance using the lens-MIMO with lower hardware complexity. More precisely, we summarize our contributions as follows.
\begin{itemize}
    \item First, we investigate the received signal of the lens-MIMO for the cooperative V2V localization and derive the CRLBs of the position and orientation of the vehicles for the given CRLB of AoA. With the considered system model, we confirm that the lens-MIMO is superior to the conventional ULA under the specific condition of the lens's design parameters, which is shown in terms of focal length and lens aperture.
    \item We propose a lens-MIMO AoA estimation algorithm with low complexity exploiting a ratio of only the two most strongly received signals at the antenna elements (R2SA).
    \item We subsequently employ the successive interference cancellation (SIC) technique to the R2SA in order to estimate multiple AoAs. It is shown that the inherent energy-focusing property of lens-MIMO helps the suppression of interference among the multiple incoming paths, especially in the intersection scenario. Consequently, this indicates that the lens-MIMO with SIC can further improve the estimation accuracy of multiple AoAs. 
    \item With the estimated AoAs, we explore the feasibility of a relative localization method that solves sensing equations (SE) associated with AoA-based geometric models, whose performance asymptotically approaches the maximum likelihood (ML) localization.
    \item By simulations, we demonstrate that the lens-MIMO considerably outperforms conventional ULA. Furthermore, in both position and orientation estimation, the performance of the proposed relative localization scheme approaches to the derived CRLB and satisfies the target requirements for 5G-V2X positioning services, particularly in high signal-to-noise ratio (SNR) regions and with a larger number of antennas.
\end{itemize}

\textbf{Notation:}
For a matrix $\mathbf{A}$, $\mathbf{A}^{\text{T}}$, $\mathbf{A}^{\text{H}}$, $\mathbf{A}^{-1}$, and $\text{Tr}(\mathbf{A})$ are the transpose, conjugate transpose, inverse, and trace operation respectively. $\mathbb{E}$ is the expectation of a random variable.
\begin{figure}
 \renewcommand{\figurename}{Fig.}
    \centering
    \includegraphics[width=0.98\linewidth]{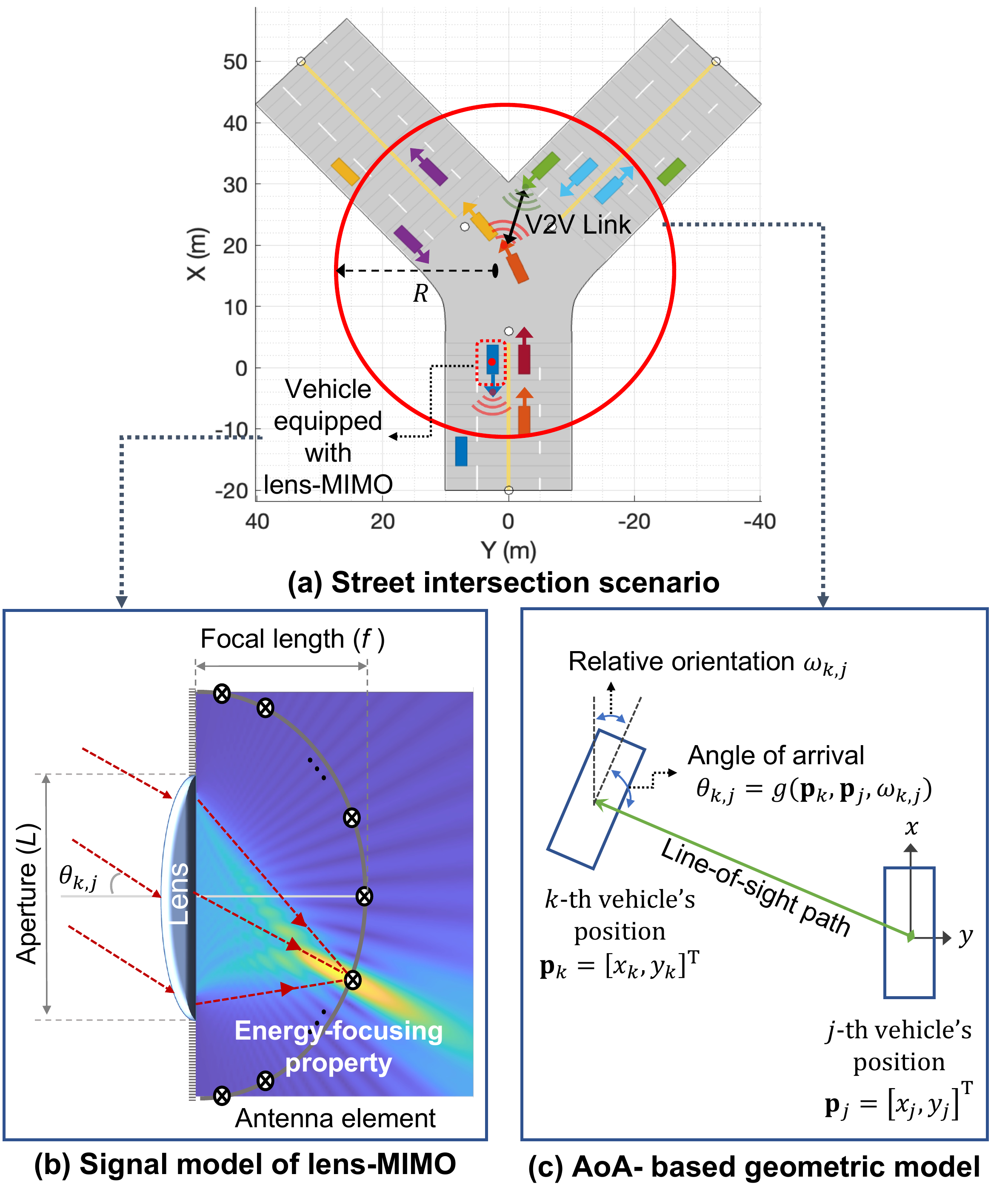}
    \caption{Cooperative V2V scenario with lens-MIMO.}
    \label{f1}
\end{figure}
\section{system model}\label{sec2}
In this section, we comprehensively describe the received signal model of lens-MIMO systems in terms of wave optics on the geometrical model in the street intersection, where each road has different directions, as shown in Fig. \ref{f1}(a). Specifically, each vehicle in a set of vehicles $\mathcal{V}=\{1,\dots,N_{v}\}$ is approaching or crossing the intersection. Their communication range is assumed to be $R$ in two-dimensional space. Fig. \ref{f1}(b) shows that each vehicle is equipped with lens-MIMO, which possesses an energy-focusing characteristic that facilitates simplified AoA estimation using a subset of the array. Fig. \ref{f1}(c) illustrates that the position of the $k$-th vehicle is $\mathbf{p}_k=\left[x_k, y_k\right]^\text{T}\in\mathbb{R}^2$ with its relative orientation $\omega_{k,j}\in[0, 2\pi)$ along with the $X$-axis while the $j$-th connected vehicle has a position of $\mathbf{p}_j$. Further, we assume that the $j$-th vehicle is known to be aligned with the X-axis, but we can assume any angle because it does not affect the estimation of the relative orientation. 

\subsection{Signal model}
We consider a mmWave lens-MIMO system, as depicted in Fig. \ref{f1}(b), where vehicles are equipped with $N$ received antenna elements. Precisely, the received signal at the $n$-th antenna element of the $k$-th vehicle can be defined by an AoA $\theta_{k,j}$ from the $j$-th to $k$-th vehicle as follows \cite{b6}:
\begin{equation}
    [\mathbf{y}_{k, j}]_{n}
    =\frac{h_{k,j}L}{\sqrt{\rho_{\text{o}}^{k,j}x}}\,\text{sinc}\left[\frac{L}{\lambda}\left\{\sin{\theta_{n}}-\sin{\theta_{k,j}}\right\}\right]e^{-j\Phi_{0}}+[\mathbf{n}]_{n},\label{eq1}
\end{equation}
where $h_{k,j}$ and $\rho_{\text{o}}^{k,j}$ are the complex channel gain and path-loss coefficient between the $k$-th and $j$-th vehicles, respectively; and $L$, $\lambda$, and $x$ are a lens aperture, the wavelength of operating frequency, and the distance from the rear surface of the lens to the antenna array, respectively. The constant phase term, which is represented as $\Phi_{0}=2\pi x/\lambda$, depends only on the $x$. Further, the antenna elements on the arc array are placed at the respective focal points corresponding to each angle of $\theta_{n}\in\left\{\sin^{-1}\left(\frac{d_{\text{lens}}}{x}n\right):n=-\frac{N-1}{2},\dots,\frac{N-1}{2}\right\}$, where $d_{\text{lens}}$ represents an antenna spacing, which is equal to $f/L$. Additionally, $\mathbf{n}\in\mathbb{C}^{N\times1}$ is the additive white Gaussian noise (AWGN) vector following the distribution of $\mathcal{CN}(0, \sigma^{2}\mathbf{I}_{N})$.

It can be seen in \eqref{eq1} that its amplitude depends on the AoA, while its received phase is not affected
by AoA but depends only on the distance from the lens to the antenna array.

\subsection{Geometric model of V2V use cases}
We consider the sensing-based scheduling employed in 3GPP V2V unicast communication. In this scheme, each target link is assigned to a separate sub-channel \cite{interf1, interf2}. Based on the geometric structure illustrated in Fig. \ref{f1}(c), the measurement model for the AoA of the line-of-sight (LoS) path from the $j$-th to the $k$-th vehicle can be expressed as
\begin{align}
\tilde{\theta}_{k,j} = g(\mathbf{p}_{k},\mathbf{p}_{j}, \omega_{k,j}) + n(\theta_{k,j}),\label{eq2}
\end{align}
where $\tilde{\theta}_{k,j}$ and $\omega_{k,j}\forall k, j\in\mathcal{V}$ are the measured AoA at the $k$-th vehicle and relative orientation of the $k$-th vehicle with respect to the $j$-th vehicle's orientation, respectively. While the measurement noise model in \eqref{eq2} can be different depending on how to sense the incident angles, we assume that the measured AoA's noise $n(\theta_{k,j})$ follows a Gaussian distribution with zero mean and variance $\sigma_{n}^{2}(\theta_{k,j})$, which is widely adopted in \cite{loc1,loc2}. Recalling sensing-based scheduling, it allows the measured AoAs to be independent and identically distributed (i.i.d.) as formulated in \eqref{eq2}. The actual AoA of the LoS path from the $j$-th vehicle to the $k$-th vehicle, denoted as $g(\cdot)$, can be expressed in terms of the position and orientation parameters as follows:
\begin{equation}
g(\mathbf{p}_{k},\mathbf{p}_{j},\omega_{k,j})=\tan^{-1}\left( \frac{y_j-y_k}{x_j-x_k} \right) - \omega_{k,j}.\label{eq5}
\end{equation}

The model in \eqref{eq5} depends on the position ($\mathbf{p}_{k}$, $\mathbf{p}_{j}$) and orientation ($\omega_{k,j}$). 
While this model has the potential to result in multiple solutions for the localization parameters, we will explore the conditions that are required to ensure a unique solution in Section \ref{sec4-4}. Furthermore, we assume that the variance of AoA measurement, denoted as $\sigma_{n}^{2}(\theta_{k,j})$, has a lower bound which is equal to the CRLB of the AoA $\theta_{k,j}$. In the cases of conventional ULA and lens-MIMO systems, their CRLBs of AoA can be readily derived as follows \cite{b6}:
\begin{align}
    \text{CRLB}&_{\text{ULA}}(\theta)=\frac{6\sigma^2}{N (N^2-1) d_{\text{ULA}}^2 \cos^{2}{\theta}},\label{eq3}\\
\begin{split}    
    \text{CRLB}&_{\text{Lens}}(\theta)=\frac{\sigma^2}{2}\sum_{n}\mathbf{a}_{n}^{2}(\theta)\Bigg/ \Bigg[ \left\{\sum_{n}\mathbf{a}^{2}_{n}(\theta)\right\}\\
    \times&\sum_{n} \left\{ \frac{\partial}{\partial\theta} \mathbf{a}_{n}(\theta)\right\} - \left\{\sum_{n}\mathbf{a}_{n}(\theta)\frac{\partial}{\delta\partial} \mathbf{a}_{n}(\theta)\right\}^2 \Bigg],\label{eq4}    
\end{split}
\end{align}
where $d_{\text{ULA}}$ is an antenna spacing of ULA, and $\mathbf{a}_{n}(\theta)=\frac{L}{\sqrt{x}}\text{sinc}\left[\frac{L}{\lambda}\left\{\sin{\theta_{n}}-\sin{\theta}\right\}\right]$ is a steering vector whose elements are the amplitude of the received signal in \eqref{eq1}. As the distance $x$ between the antenna array and the rear of the lens increases, it leads to a reduction in received amplitude, resulting in potential inaccuracies in AoA estimation.

\section{Theoretical bound for localization}\label{sec3}
In this section, we investigate the CRLBs for AoA estimation in both the lens-MIMO and ULA. Furthermore, we derive the theoretical lower bound for estimation of position and orientation.
\subsection{CRLB Analysis for AoA Estimation}\label{3-2}
Accurate estimation of the AoA is crucial for achieving precise positioning. 
Now, we compare the CRLBs of the AoA for the lens-MIMO and ULA. However, the direct comparison of (4) and (5) is challenging to properly explain because the AoA $\theta$ in \eqref{eq4} is involved squared amplitude in a non-linear fashion, thus, we require a more insightful form of \eqref{eq4} to make it easier to compare both CRLBs.

Hence, we assume that all antenna elements of lens-MIMO are placed on the focal region with critical antenna spacing. In other words, the distance from the rear of the lens to the array is the focal length (i.e., $x=f$), and the antenna placement can be represented by the critical angular set, defined by $\theta_{n}\in\mathcal{S}_{\theta_{n}}=\left\{\sin^{-1}\left(\frac{d_{\text{lens}}}{f}n\right):n=-\frac{N-1}{2},\dots,\frac{N-1}{2}\right\}$ with $d_{\text{lens}}=\frac{f}{L}\lambda$ given by in \cite{b7}. Now, to obtain a derivative of the amplitude in \eqref{eq4}, let us define $\boldsymbol{\mu}_1$ and $\boldsymbol{\mu}_2$ as
\begin{equation}
\left[\boldsymbol{\mu}_1\right]_{n}=\text{sinc}\left(\frac{Ld_{\text{lens}}n}{f\lambda}-\frac{L}{\lambda}\sin{\theta}\right),\label{eq19}
\end{equation}
\begin{equation}
\left[\boldsymbol{\mu}_2\right]_{n}=\frac{\partial}{\partial\theta}\text{sinc}\left(\frac{Ld_{\text{lens}}n}{f\lambda}-\frac{L}{\lambda}\sin{\theta}\right).\label{eq20}
\end{equation}

It may be readily denoted that $\mathbf{a}_{n}=\frac{L}{\sqrt{f}}\boldsymbol{\mu}_1$ and $\frac{\partial}{\partial\theta}\mathbf{a}_{n}=\frac{L^2}{\lambda\sqrt{f}} \cos{\theta} \boldsymbol{\mu}_2$. Then, the CRLB of the lens-MIMO in \eqref{eq4} can be represented by
\begin{equation}
\begin{split}
    \text{CRLB}&_{\text{Lens}}(\theta)\\
    &=\frac{f\lambda^2\sigma^2}{2p^2 L^4 \cos{\theta}^2}\cdot\frac{\boldsymbol{\mu}_1^\text{T} \boldsymbol{\mu}_1}{[\boldsymbol{\mu}_1^\text{T} \boldsymbol{\mu}_1][\boldsymbol{\mu}_2^\text{T} \boldsymbol{\mu}_2]-[\boldsymbol{\mu}_1^\text{T} \boldsymbol{\mu}_2]}.\label{eq21}    
\end{split}
\end{equation}

By the property of the lens-MIMO, which works like a discrete Fourier transform (DFT) beamformer, we subsequently obtain the inequalities as follows:
\begin{equation}
    \begin{split}
        &\boldsymbol{\mu}_{1}^\text{T}\boldsymbol{\mu}_{1}=1,\\
        &\boldsymbol{\mu}_{1}^\text{T}\boldsymbol{\mu}_{2}=0,\\
        1&\leq\boldsymbol{\mu}_{2}^\text{T}\boldsymbol{\mu}_{2}\leq2.\label{eq22}
    \end{split}
\end{equation}

By these inequalities, the CRLB of lens-MIMO can be simplified as follows:
\begin{equation}
    \text{CRLB}_{\text{Lens}}(\theta)=\frac{f\lambda^2\sigma^2}{2p^2 L^4 \cos{\theta}^2}\cdot\frac{1}{[\boldsymbol{\mu}_2^\text{T} \boldsymbol{\mu}_2]}.\label{eq23}
\end{equation}

To ensure that $\text{CRLB}_{\text{Lens}} \leq \text{CRLB}_{\text{ULA}}$, the condition, in which the performance of the lens-MIMO would prevail over that of ULA, is readily derived in terms of the lens aperture and focal length as follows:
\begin{equation}
    \text{CRLB}_{\text{Lens}} \leq \text{CRLB}_{\text{ULA}}\,\text{where}\,f\leq\frac{12L^3}{(2L+1)(L+1)\lambda^4}.\label{eq24}
\end{equation}
$Proof$: See Appendix A.

Suppose that the focal length is set to the minimum $f=\frac{L}{2}$, which is proven in \cite{b8}, then the condition in \eqref{eq24} can be expressed in terms of lens aperture $L$ as follows:
\begin{equation}
\frac{12L^3}{(2L+1)(L+1)\lambda^4}>\frac{L}{2}.\label{eq25}
\end{equation}

Given that LTE-V2X and NR-V2X operate at frequencies in the GHz range, it follows that the wavelength $\lambda$ is much smaller than 1 meter (e.g., $\lambda$ = 0.1~m and 0.01~m for frequencies of 2.4~GHz and 28~GHz, respectively). Consequently, \eqref{eq25} can be readily proved to hold for $L=k\lambda$, where $k$ is an integer and is practically greater than 10 in order to efficiently concentrate the received signal energy on the lens's surface \cite{b4, b6, b8}. For given system parameters of the operating frequencies in the GHz spectrum, the inequality is always satisfied within the feasible region of the lens's aperture size. When we consider values of $\lambda$ equal to 0.1~m and 0.01~m, we find that \eqref{eq25} is satisfied if the value of $k$ is greater than $10^{-2}$ and $10^{-3}$ respectively. However, the lens aperture that contradicts these conditions (i.e. $L<10^{-3}~\text{m}$ and $L<10^{-5}~\text{m}$ for frequencies of 2.4~GHz and 28~GHz, respectively) is impractical for the lens-MIMO system. It can be noted that the AoA's CRLB of the lens-MIMO with a feasible aperture size is always less than that of the ULA in NR-V2X communication systems.

In \eqref{eq24} and \eqref{eq25}, we highlighted that the lens, whose focal length is half of the lens' aperture, has advantages of performance and implementation of a small form factor of the lens-MIMO. However, its lens gets thicker for higher permittivity and could suffer higher lens transmission loss \cite{lens}. Meanwhile, as the focal length increases, the amplitude gain at the antenna array could get a little more path loss between the rear of the lens and antenna array as in \eqref{eq4}. Therefore, we could also consider a trade-off between the lens' transmission and path loss in choosing the focal length, but this is beyond our scope and would be considered in future research.

\begin{figure}
 \renewcommand{\figurename}{Fig.}
\centering
{\includegraphics[width=0.99\linewidth]{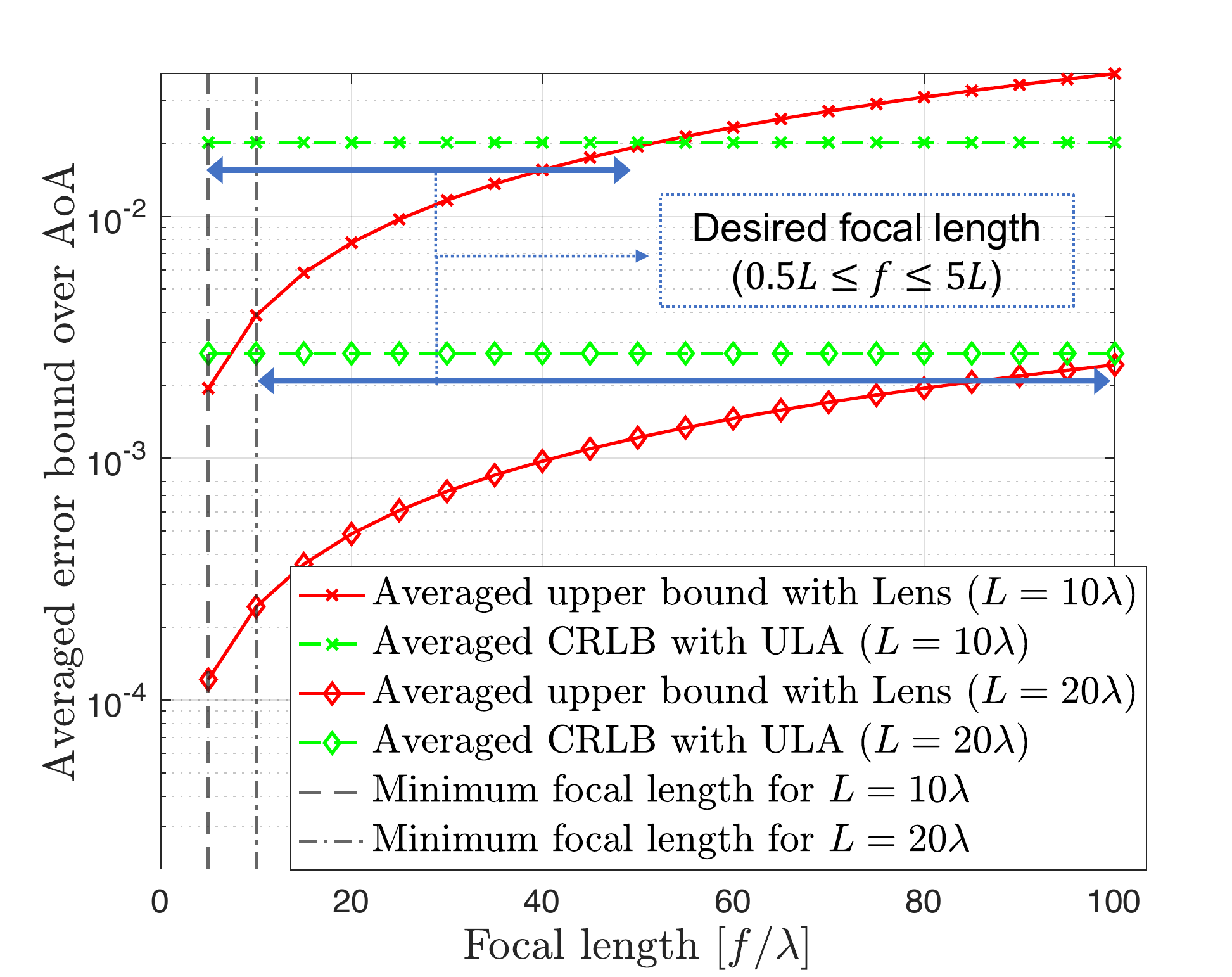}}
\caption{Error bound as a function of the focal length.}
\label{f3}
\end{figure}
Subsequently, Fig. \ref{f3} compares the upper bound of lens-MIMO's CRLB, which is derived as \eqref{b5} in Appendix A, and the conventional ULA in \eqref{eq3} as the focal length increases for a given lens aperture. For a fair comparison with the same antenna elements, it is recalled that we placed the ($2L+1$) antennas for both the lens-MIMO and ULA, where the signal-to-noise ratio (SNR) is assumed to be 5~dB and antenna spacing is $f/L$ for lens-MIMO and $\lambda/2$ for ULA, respectively. Further each $Y$-axis and $X$-axis in Fig. \ref{f3} represent the CRLB averaged over AoA $\theta\in[-\frac{\pi}{2},\frac{\pi}{2}]$ and focal length $f$ of wavelength unit. For both cases of the $10\lambda$ and $20\lambda$ lens apertures, the lens-MIMO performs much better at the minimum focal length (i.e., $f=5\lambda$ for $L=10\lambda$ and $f=10\lambda$ for $L=20\lambda$), and this performance is sustained until the focal length is about five times the lens aperture (i.e., $f=50\lambda$ for $L=10\lambda$ and $f=100\lambda$ for $L=20\lambda$). Fig. \ref{f3} confirmed that lens-MIMO is superior to the conventional ULA for a certain condition of focal length, which is from the minimum distance $L/2$ to five times of lens' aperture $5L$. It should be noted that the suitable focal length design for the configuration of the lens' aperture is essential to enhance the positioning and orientation estimation performance.

It is important to highlight that the lens-MIMO can achieve higher accuracy in estimating the position and orientation, given the specific design parameters of the lens, in comparison to the ULA.

\subsection{CRLB derivation for position \& orientation}
We now derive the theoretical bounds of the position and orientation estimates in order to compare the localization performance. Initially, we assume that all vehicles communicate without interference by assigning each pair of vehicles to different sub-channels in terms of the sensing-based scheduling of 3GPP V2V communication. We also consider a sparse mmWave channel with a single LoS path, which is assumed to be a prevalent link state in V2V unicast channels with short distances \cite{c14-1,losp1,losp2}. We will provide further discussion on this topic in Section IV.
Let $\mathcal{V}_k$ be the set of neighboring vehicles within the communication range of the $k$-th vehicle. Assuming that AoAs in \eqref{eq2} are independent and identically distributed (i.i.d.) Gaussian random variables, the joint probability density function of a vector of AoAs $\boldsymbol{\theta}$ is given by 
\begin{equation}
\begin{split}
f(\boldsymbol{\theta}|\mathbf{p},\boldsymbol{\omega})& = \prod_{k\in\mathcal{V}}\prod_{j\in\mathcal{V}_k} f(\tilde{\theta}_{k,j}|\mathbf{p}_{k},\mathbf{p}_{j},\omega_{k,j})\\
& = \prod_{k\in\mathcal{V}}\prod_{j\in\mathcal{V}_k}\frac{1}{ \sqrt{2\pi\sigma_{n_{k,j}}^{2}}}\text{exp}\left(- \frac{(\tilde{\theta}_{k,j}-\alpha_{k,j})^{2}} {2\sigma_{n_{k,j}}^{2}} \right),\label{eq6}
\end{split}
\end{equation}
where $\boldsymbol{\theta}$, $\mathbf{p}$, and $\boldsymbol{\omega}$ are vectors consisting of all AoA measurements, position, and orientation, respectively; $\alpha_{k,j}$ is an abbreviated form of the $g(\mathbf{p}_k,\mathbf{p}_j,\omega_{k,j})$ in \eqref{eq5}. The variance of the AoA measurement noise, denoted as $\sigma_{n_{k,j}}^{2}$, depends on the AoA of the LoS path between the $j$-th and $k$-th vehicles, as noted in \eqref{eq4}.

To investigate the lower bounds of the position and orientation accuracy, we define a vector consisting of unknown location parameters as
\begin{equation}    \boldsymbol{\eta}=\left[\boldsymbol{\eta}_{1}^{\text{T}},\dots,\boldsymbol{\eta}_{N_v}^{\text{T}} \right]^\text{T}\in\mathbb{R}^{3N_{v}\times1},\label{eq7}
\end{equation}
in which $\boldsymbol{\eta}_{k}^{\text{T}}$ consists of the unknown parameters (position $\mathbf{p}_k$ and orientation $\omega_{k,j}$) for the $k$-th vehicle. Defining $\hat{\boldsymbol{\eta}}$ as the unbiased estimator of $\boldsymbol{\eta}$, the error variance satisfies the inequality as \cite{b9},
\begin{equation}
\mathbb{E}_{\boldsymbol{\theta}|\boldsymbol{\eta}}[(\boldsymbol{\eta}-\hat{\boldsymbol{\eta}})(\boldsymbol{\eta}-\hat{\boldsymbol{\eta}})^{\text{H}}]\geq F^{-1}(\boldsymbol{\eta}),\label{eq8}
\end{equation}
where $\mathbb{E}_{\boldsymbol{\theta}|\boldsymbol{\eta}}[.]$ denotes the expectation parameterized by the unknown parameter $\boldsymbol{\eta}$, and the Fisher information matrix (FIM) $F(\boldsymbol{\eta})$ is defined by 
\begin{equation}
F(\boldsymbol{\eta})=-\mathbb{E}_{\boldsymbol{\theta}|\boldsymbol{\eta}}\left[\frac{\partial^{2}\ln{f(\boldsymbol{\theta}|\boldsymbol{\eta})}}{\partial\boldsymbol{\eta}\partial\boldsymbol{\eta}^{\text{T}}}\right].\label{eq9}
\end{equation}

The $3N_v\times3N_v$ FIM $F(\boldsymbol{\eta})$ may be re-constructed in blocks as an $N_v \times N_v$ sub-block matrix for each unknown parameter. It is formed as
\begin{align}
F(\boldsymbol{\eta}) 
=\left [
\begin{array}{c}
F_{\mathbf{xx}} \quad F_{\mathbf{xy}} \quad F_{\mathbf{x\boldsymbol{\omega}}}\\
F_{\mathbf{yx}} \quad F_{\mathbf{yy}} \quad F_{\mathbf{y\boldsymbol{\omega}}}\\
F_{\mathbf{\boldsymbol{\omega}x}} \quad F_{\mathbf{\boldsymbol{\omega}y}} \quad F_{\mathbf{\boldsymbol{\omega}\boldsymbol{\omega}}}\\
\end{array}
\right].\label{eq10}
\end{align}

The elements of $F_{\mathbf{xx}}\in\mathbb{C}^{N_v \times N_v}$ matrix may be readily obtained by the geometric model in \eqref{eq5} as follows:
\begin{equation}
\begin{split}
    [F&_{\mathbf{xx}}]_{i,j}=\begin{cases}
    \sum_{j\in\mathcal{V}_k}\frac{1}{ \sqrt{2\pi}\sigma_{n_{k,j}}^{3}}\frac{(y_j-y_k)^2}{d^4_{j,k}}, & \text{for}\, i=j\\
    0, & \text{otherwise},\label{eq11}
    \end{cases}  
\end{split}
\end{equation}
where $d_{j,k}$ is the distance from the $k$-th to $j$-th vehicles. The details are proved in Appendix B.

The diagonal term of each sub-matrix may also be readily derived as
\begin{align}
[F_{\mathbf{yy}}]_{i,i}&=\sum_{j\in\mathcal{V}_k}\frac{1}{\sqrt{2\pi}\sigma_{n_{k,j}}^{3}}\frac{(x_j-x_k)^2}{d^4_{j,k}},\label{eq12}\\
[F_{\mathbf{xy}}]_{i,i}&=[F_{\mathbf{yx}}]_{i,i}\notag\\
&=\sum_{j\in\mathcal{V}_k}\frac{1}{\sqrt{2\pi}\sigma_{n_{k,j}}^{3}}\frac{(x_j-x_k)(y_j-y_k)}{d^4_{j,k}},\label{eq13}
\end{align}
\begin{align}
[F_{\mathbf{xw}}]_{i,i}&=[F_{\mathbf{wx}}]_{i,i}\notag\\
&=-\sum_{j\in\mathcal{V}_k}\frac{1}{\sqrt{2\pi}\sigma_{n_{k,j}}^{3}}\frac{(y_j-y_k)}{d^2_{j,k}},\label{eq14}\\
[F_{\mathbf{yw}}]_{i,i}&=[F_{\mathbf{wy}}]_{i,i}\notag\\
&=-\sum_{j\in\mathcal{V}_k}\frac{1}{\sqrt{2\pi}\sigma_{n_{k,j}}^{3}}\frac{(x_j-x_k)}{d^2_{j,k}},\label{eq15}\\
[F_{\mathbf{ww}}]_{i,i}&=\sum_{j\in\mathcal{V}_k}\frac{1}{\sqrt{2\pi}\sigma_{n_{k,j}}^{3}},\label{eq16}
\end{align}
and the rest of the entries (i.e., the off-diagonal terms of each sub-block) are zero.

By exploiting the derived FIM, we can obtain the vehicle's position and orientation error bounds, denoted by $\text{PEB}_p$ and $\text{PEB}_{\omega}$, as follows:
\begin{align}
\text{PEB}_p & \geq \sqrt{\text{Tr}\left\{\left[F(\boldsymbol{\eta})^{-1}\right]_{1:2N_{v},1:2N_{v}}\right\}},\label{eq17}\\
\text{PEB}_{\omega} & \geq \sqrt{\text{Tr}\left\{\left[F(\boldsymbol{\eta}^{-1})\right]_{2N_{v}+1:3N_{v},2N_{v}+1:3N_{v}}\right\}},\label{eq18}
\end{align}
where the operation $[.]_{i:j,i:j}$ denotes the selection of sub-matrix from the $i$-th to the $j$-th entry.

It is worth noting that in the derived FIM from \eqref{eq11} to \eqref{eq16}, the diagonal elements are associated with the AoA error variance $\sigma_{n_{k,j}}^{2}$ and the specified values of positioning parameters, such as position and distance between vehicles. Additionally, we can infer that a smaller CRLB for the AoA leads to improved accuracy in estimating the position and orientation, thus the lens-MIMO can show much better performance compared to ULA, as elaborated in Section \ref{3-2}.
\section{AoA Estimation in Lens-MIMO}\label{sec4}
Now, we investigate the characteristics of the received signal at lens-MIMO and propose a lower complexity AoA estimation scheme taking full advantage of the unique features of lens-MIMO.
\subsection{Characteristics of the received signal at lens-MIMO}\label{sec4-1}
The essential characteristic of the received signal of the lens-MIMO is depicted in Fig. \ref{f1}(b). More specifically, in the case of a single AoA, most of the energies arriving on the lens surface are focused on a few antennas. We will introduce the ratio of two adjacent antenna signals and demonstrate how the AoA may be expressed by those ratios, which will be the main idea of the proposed a lower complexity technique for the AoA estimation.

As regards $\theta_{k,j}$ in \eqref{eq1}, which is the AoA of a single path, it is the closest to the $n_{k,j}$-th critical angle in $\mathcal{S}_{\theta_{n}}$, in which the antenna element collects the strongest received power. Afterward, the AoA may be expressed as $\theta_{k,j}=\sin^{-1}\left(\frac{\lambda}{L}(n_{k,j}+e_{k,j})\right)$, where $e_{k,j}$ is the error between the angular sample $\theta_{n_{k,j}}$ and the actual AoA $\theta_{k,j}$. Hence, the amplitude of the received signal at the $n$-th antenna can be represented as
\begin{equation}
    \mathbf{a}_{n}(\theta_{k,j})=\frac{L}{f}\text{sinc}\left({n} - (n_{k,j}+e_{k,j})\right),\label{eq26}
\end{equation}
where $e_{k,j}$ should be in the range $-0.5\leq e_{k,j}\leq0.5$. The ratio of signals between the $n$-th and $(n+1)$-th element, denoted by $R_{n}(\theta_{k,j})=\mathbf{a}_{n}(\theta_{k,j})/\mathbf{a}_{n+1}(\theta_{k,j})$, is reformulated by an identity of the trigonometric function, it is given by
\begin{equation}
    \begin{split}
        R_{n}(\theta_{k,j})&=\frac{\sin(\pi(n-n_{k,j}-e_{k,j}))}{\sin(\pi(n+1-n_{k,j}-e_{k,j}))}\frac{n+1-n_{k,j}-e_{k,j}}{n-n_{k,j}-e_{k,j}}\\
        &=\frac{\cos(\pi(n-n_{k,j}))}{\cos(\pi(n+1-n_{k,j}))}\frac{n+1-n_{k,j}-e_{k,j}}{n-n_{k,j}-e_{k,j}}\\
        &=-1-\frac{1}{n-n_{k,j}-e_{k,j}},\label{eq27}
    \end{split}
\end{equation}
which is represented to
\begin{equation}
n-n_{k,j}-e_{k,j}=-\frac{1}{R_{n}(\theta_{k,j})+1}.\label{eq28}
\end{equation}

By adding \eqref{eq28} for all antenna indices, the AoA $\theta_{k,j}$ can be represented as
\begin{equation}
    n_{k,j}+e_{k,j}=\left[\frac{1}{N - 1}\sum_{n=\frac{N-1}{2}}^{\frac{N -3}{2}}\frac{1}{R_{n}(\theta_{k,j})+1}\right] - \frac{1}{2}.\label{eq29}
\end{equation}

It should be noted that an estimate of AoA can be represented by a function of the ratios between the amplitudes of adjacent antennas. In the following section, we introduce an AoA estimation scheme inspired by \eqref{eq29}.

\subsection{A single target vehicle per V2X sub-channel}\label{sec4-2}
Prior to the localization for vehicles, we need to estimate the AoA, as indicated by the measurement model of AoA in \eqref{eq2}. Furthermore, it is worth mentioning that the lens-MIMO system exhibits a peak power feature, where the received signal power, arriving at specific critical angles $\mathcal{S}_{\theta_{n}}$, is concentrated on a single antenna element. When the AoA deviates from the critical angles, this peak power feature fades away because the received power spreads out over the antenna elements, which is called a power leakage problem \cite{cb1, cb2}. This leakage problem is the worst when the AoA is in the middle of the adjacent critical angles, which leads to a degradation in the AoA estimation performance. 
To improve the AoA estimation while minimizing complexity in the presence of power leakage problem, we propose an AoA estimation algorithm called R2SA (Ratio of 2 most Strong received powers at the antenna elements). This algorithm leverages the ratio of the two strongest received powers at the lens-MIMO. Precisely, the algorithm first selects the antenna of the strongest received power and its adjacent antenna. Particularly, the antenna index of the strongest received power can indicate a rough estimate of the AoA, and the received amplitude ratio of these selected two antennas would be used to enhance the rough estimate.

We assume the V2V unicast communications in 3GPP V2X standardization, where each sub-channel is allocated to each different V2V link such that we can hold the validity of the i.i.d. V2V channels among vehicles like in \eqref{eq17} and \eqref{eq18} of Section \ref{sec3}. Following this, the $k$-th vehicle is connected with the $j$-th target vehicle in $\mathcal{V}_k$ in a sub-channel, and it measures the amplitudes of the whole antenna array and determines the antenna element $n^{*}_{k,j}$ of the strongest received power. Then, we have
\begin{equation}
    n^{*}_{k,j}=\underset{{n\in\mathcal{S}_{N}}}{\arg\max}\left|\mathbf{y}_{k,j}\right|,\label{eq30}
\end{equation}
where $\mathcal{S}_{N}=[-\frac{N-1}{2},\frac{N-1}{2}]\in\mathbb{R}^{N\times1}$ is a set of antenna elements. By setting $n_{k,j}$ to $n^{*}_{k,j}$ in \eqref{eq28}, we can estimate the error term $e_{k,j}$. Next, we propose the R2SA for the AoA estimation as follows:
\begin{equation}
\begin{split}
    \hat{\theta}&_{k,j}=\\
    &\begin{cases}
    \sin^{-1}\left(\frac{\lambda}{L}(n^{*}_{k,j}+e_{k,j})\right),&\text{for}\, -1\leq \frac{\lambda}{L}(n^{*}_{k,j}+e_{k,j}) \leq 1\\
    \sin^{-1}\left(\frac{\lambda}{L}(n^{*}_{k,j})\right),&\text{otherwise},\label{eq31}
    \end{cases}    
\end{split}
\end{equation}
where $e_{k,j}=1/(R_{n^{*}_{k,j}}(\theta_{k,j})+1)$ computed by \eqref{eq28} enhances a rough estimate of the AoA. It should be noted that \eqref{eq31} without adjusting $e_{k,j}$ is a rough estimate of the AoA, called maximum antenna selection (MS). Particularly, the rough estimate of MS is limited by the angular resolution confined by the total number of antennas. The proposed R2SA algorithm in \eqref{eq31} does not require the correlation process or exhaustive searching by exploiting only two adjacent antennas.

\subsection{Multiple target vehicles per V2X sub-channel}\label{sec4-3}
We now assume that the $k$-th vehicle can be simultaneously connected with multiple vehicles within the communication range in the same sub-channel to improve spectral efficiency and latency. It is possible to assume that the $k$-th vehicle can receive positioning reference signals from multiple target vehicles for localization purposes. However, an issue arises regarding the scheduling of simultaneous connections of the multiple target vehicles on the same sub-channel. For this issue, we expect that collisions are more likely to occur at the front and rear of the subject vehicle within its proximity region. Hence, when moving into the intersection, we can define high-priority vehicles as those moving toward us in the opposite lane or crossroads. However, the specific implementation of scheduling is beyond the scope of this paper. Specifically, we will look into the performance of the R2SA for the multiple AoAs estimation in the V2V environment, where the the $k$-th vehicle is assumed to be capable of suppressing the interference of multiple vehicles through the successive interference cancellation (SIC) method. Since the signals of different incident angles are focused on different focal points at the antenna array, they tend to be physically separated over different antenna elements. As a result, this leads to the inter-path interference suppression, and helps the SIC processing to further improve the multiple AoAs estimation.
\begin{algorithm}[!t]
\caption{AoA Estimation algorithm}\label{alg}
\KwInput{\\
Received signal at $k$-th vehicle, $\tilde{\mathbf{y}}_{k}^{1}=\mathbf{y}_k$.\\
Signal for cancellation, $\tilde{\mathbf{y}}_{k}^{0}=0$.}
{\For{$j\gets1$ \KwTo $\vert\mathcal{V}_{k}\vert$}{Set the eliminated signal $\tilde{\mathbf{y}}_{k}^{j}=\tilde{\mathbf{y}}_{k}^{j}-\tilde{\mathbf{y}}_{k}^{j-1}$.\\
Select the antenna index using \eqref{eq30}.\\
{\SetKwProg{Fn}{Stage 1}{}{}
\Fn{Ratio calculation}{
    {Measure $\left|[\tilde{\mathbf{y}}_{k}^{j}]_{n^{*}_{k,j}}\right|$ and $\left|[\tilde{\mathbf{y}}_{k}^{j}]_{(n^{*}_{k,j}+1)}\right|$.\\
    Calculate $R_{n^{*}_{k,j}}$ using \eqref{eq27} and obtain $e_{k,j}=1/\left(R_{n^{*}_{k,j}}+1\right)$.
    }}}
\SetKwProg{Fn}{Stage 2}{}{}
\Fn{Fine-tuning of AoA}{
    \uIf{$-1\leq \frac{\lambda}{D}(n^{*}_{k,j}+e_{k,j}) \leq 1$}{
    $\hat{\theta}_{k,j}=\sin^{-1}\left(\frac{\lambda}{D}\left(n^{*}_{k,j}+e_{k,j}\right)\right)$,}
    \Else{
    $\hat{\theta}_{k,j}=\sin^{-1}\left(\frac{\lambda}{D}\left(n^{*}_{k,j}\right)\right)$.}}
    Calculate SIC signal using \eqref{eq32}.}}
\KwOutput{$\hat{\boldsymbol{\theta}}_k=[\hat{\mathbf{\theta}}_{k,1},\dots,\hat{\mathbf{\theta}}_{k,J}]$.}
\end{algorithm}

Further, to estimate all the AoAs $\theta_{k,j}\forall j\in\mathcal{V}_k$ in a sub-channel, the antenna index of the strongest received power at the $k$-th vehicle is first selected as in \eqref{eq30}, and the selected antenna is used to estimate the first AoA in terms of a single AoA estimation scheme in \eqref{eq31}. Next, the SIC subtracts the estimated first strongest path from the received signal. Subsequently, we select the other antenna index of the strongest received power at each SIC step and run the R2SA. The SIC continues until the AoA estimation of all the target vehicles is finished. For each iteration of SIC, the signal for estimating the $j$-th AoA can be defined as follows:
\begin{equation}
    \tilde{\mathbf{y}}_{k}^{j}=\tilde{\mathbf{y}}_{k}^{j-1}- 
    \tilde{\mathbf{y}}_{k}^{j-1}\circ\bar{\mathbf{a}}(\hat{\theta}_{k,j-1}),\label{eq32}
\end{equation}
where $[\bar{\mathbf{a}}(\hat{\theta}_{k,j-1})]_{n}=\frac{L}{\sqrt{f}}\text{sinc}\left(\frac{L}{\lambda}(\sin\theta_{n}-\sin\hat{\theta}_{k,j-1})\right)$ is a known pattern of the lens-MIMO's received signal at the estimated AoA $\hat{\theta}_{k,j-1}$ in \eqref{eq4}. The signal in \eqref{eq32} is the input to the next SIC iteration. Based on \eqref{eq32}, we select the antenna index of the strongest received power for the $j$-th AoA estimation. This way, each AoA from different vehicles is estimated sequentially at every stage of SIC. We repeat the process of AoA estimation until we finish estimating all the AoAs. Algorithm 1 summarizes the process.
\subsection{Analysis of R2SA}\label{per}
\begin{figure}[t]
 \renewcommand{\figurename}{Fig.}
     \centering
     \begin{subfigure}[b]{0.98\linewidth}
         \centering
         \includegraphics[width=\linewidth]{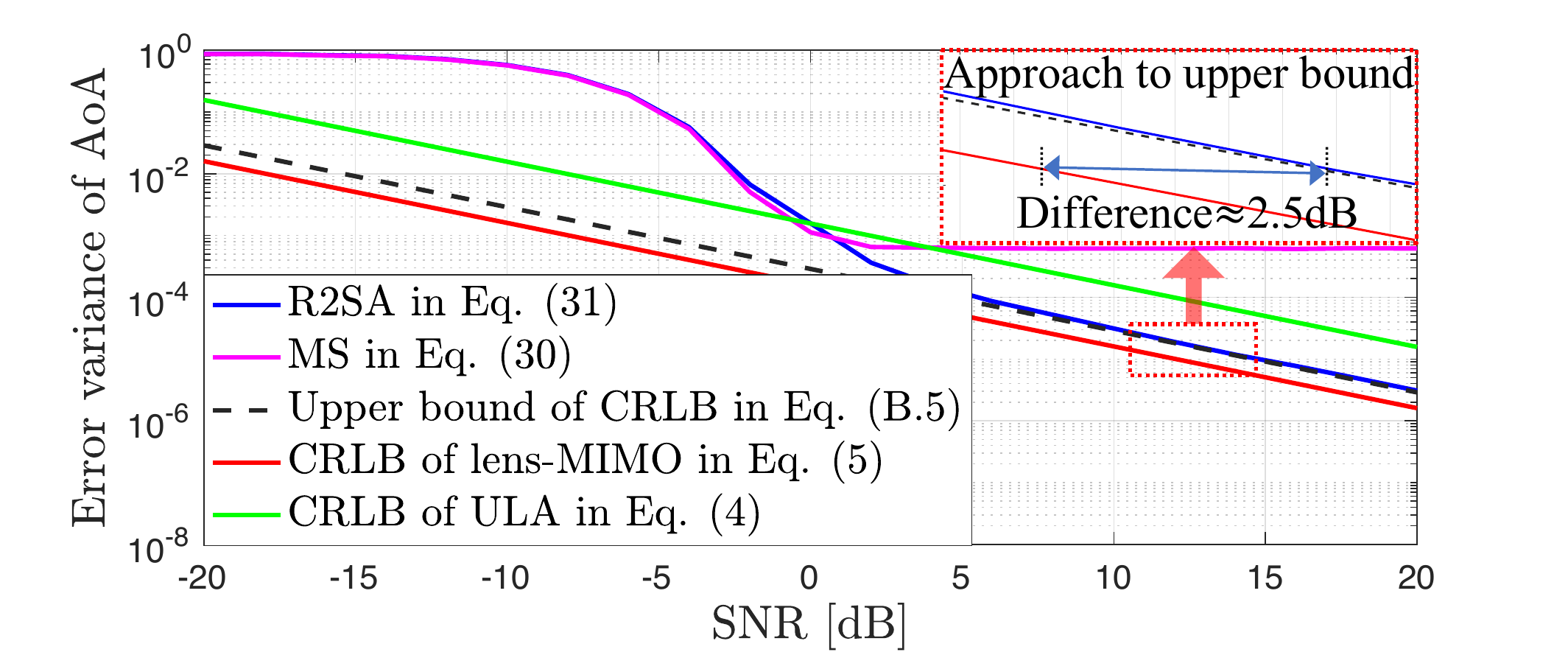}
         \caption{Single target}
     \end{subfigure}
        
     \begin{subfigure}[b]{0.98\linewidth}
         \centering
         \includegraphics[width=\linewidth]{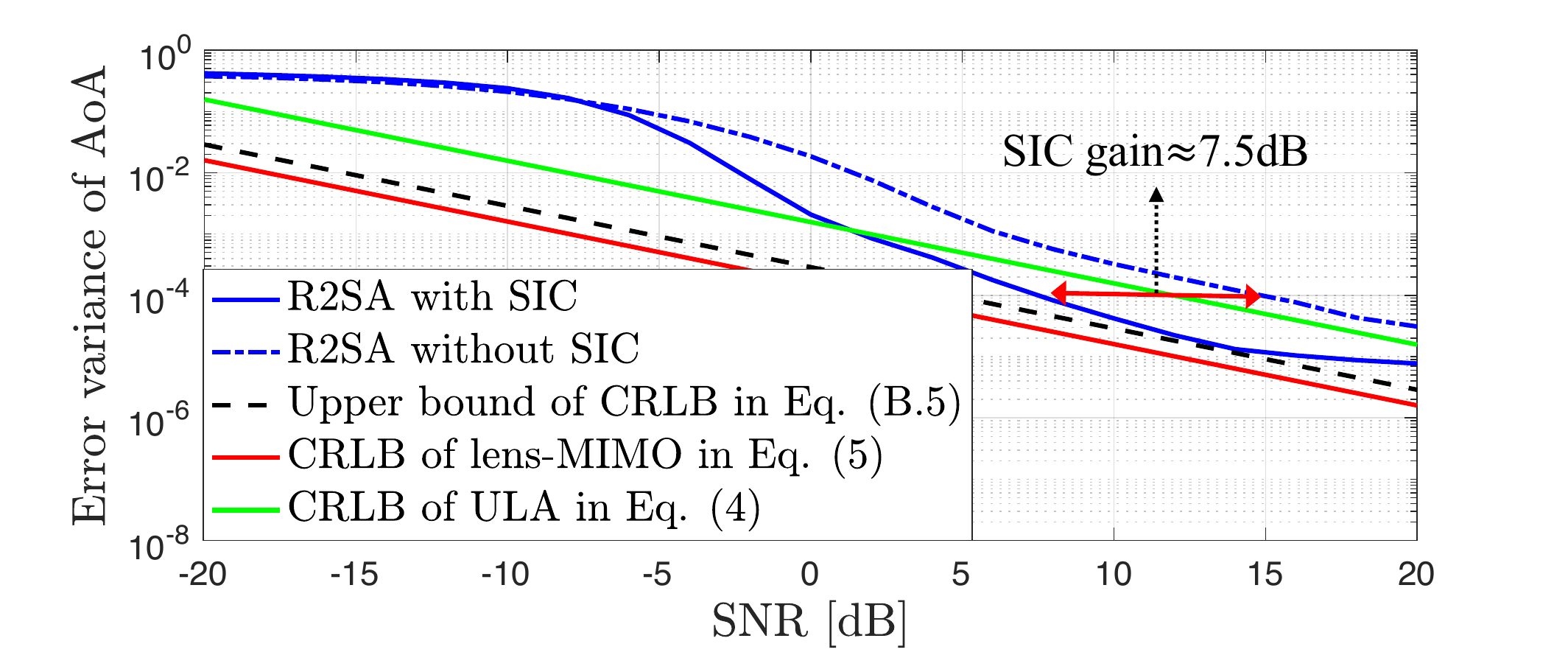}
         \caption{Multi-target with spatially separated signals}
     \end{subfigure}     
     \begin{subfigure}[b]{0.98\linewidth}
         \centering
         \includegraphics[width=\linewidth]{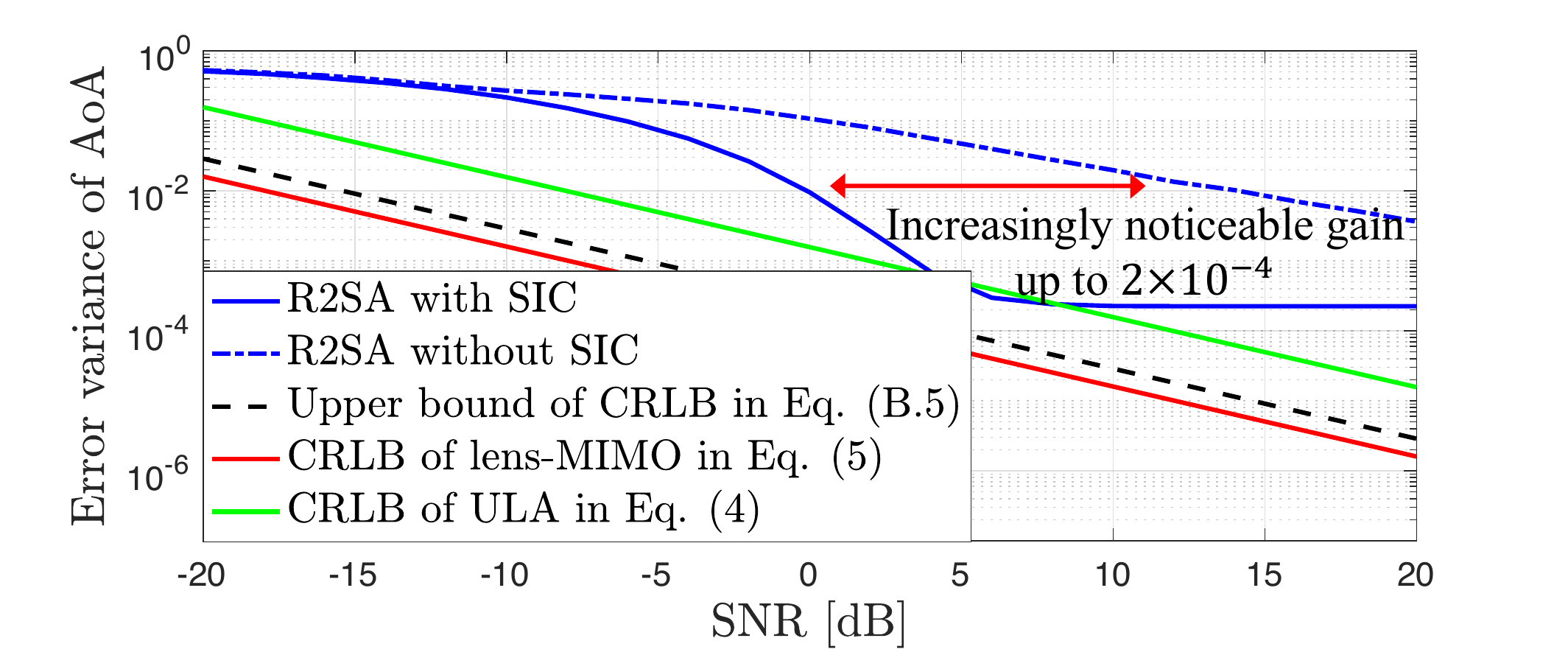}
         \caption{Multi-target with spatially correlated signals}
     \end{subfigure}
        \caption{Error variance versus SNR for a comparison in different scenarios.}
        \label{f4}
\end{figure}
\subsubsection{Performance}We first evaluate the R2SA for a different number of target AoAs and compare it with its CRLBs as the signal-to-noise ratio (SNR) increases. Fig. \ref{f4} shows the error variances of the R2SA estimate with and without SIC for three cases: (a) a single target vehicle per sub-channel, (b) two paths of two target vehicles in the same sub-channel, which are angularly separated by more than lens-MIMO resolution, (c) two paths in the same sub-channel within the lens-MIMO resolution, where the lens's resolution $\frac{1}{N}$ is $\frac{1}{2L+1}$, and the number of antennas is set to 31. The difference between (b) and (c) is whether or not two paths are so close that two paths end up with the same antenna index of the strongest received power.

On a single AoA estimation in (a), we can see that the difference between the lens's CRLB in \eqref{eq5} and its derived upper bound in \eqref{b5} is about 2.5~dB for the same performance. Additionally, R2SA converges to the upper bound of the lens-MIMO at about 7.5~dB, but MS in \eqref{eq30} is limited by the lens-MIMO resolution. The R2SA with SIC in (b) further improves the performance by $7.5$~dB at $10^{-4}$. Besides, its SIC gain in (c), where two paths focus on the same antenna, is most pronounced at $2\times10^{-4}$ error variance because the ratio in \eqref{eq27} is determined by the same antenna index, which results in a great deal of interference for the R2SA without SIC. It should be noted that the R2SA without SIC in (b) is better than the those in (c) because the inherent capability of the lens-MIMO to focus energy at different focal points allows for the physical suppression of interference from two paths with different AoAs. However, in both cases (b) and (c), an error floor is observed in the high SNR region due to the residual interference of SIC, while the case of (c) shows early error floor due to the lack of the physical separation of the two paths energies over the array elements. To overcome the error floor, a larger number of antenna elements is required and will be addressed in Section~\ref{sec5}.

In the use case of the street intersection depicted in Fig. \ref{f1}, we think that the (b) case happens more often than the (a) case if the V2X MAC scheduler connects those of vehicles on the opposite side of the lane and crossroad lane with higher priority. Thus, we statistically evaluate a probability that the difference between the two angles is greater than the resolution of lens-MIMO, where vehicles are dropped on the road as a Poisson point process (PPP) with different vehicle densities (i.e., the expected value of the number of vehicles per unit area). For computational convenience, we introduce a sign function, which is represented as $\text{sgn}(x)$, and it has 1 when $x\geq0$ and 0 otherwise. In addition, we define the angle difference as $\theta_{k}^{j,i}=|\theta_{k,j}-\theta_{k,i}|\in\mathbb{R}^{N_{v}(N_{v}-1)}$, then the probability that the AoAs of two paths are separated by more than $1/N$ is defined as
\begin{table}[!t]
\caption{\label{T1}Separation probability, $p_{\text{sep}}$}
\centering
\begin{tabular}{m{8em}|m{5.5em} m{5.5em} m{5.5em}}
\hline
\multirow{2}{10em}{Vehicle density (cars/km/lane)} & \multicolumn{3}{c}{The number of antennas, $N$} \\
& 15 & 31 & 61   \\ 
\hline\hline
10 & 0.7513 & 0.8749 & 0.9355 \\
\hline
20 & 0.7493 & 0.8650 & 0.9304 \\
\hline
40 & 0.7311 & 0.8544 & 0.9251 \\
\hline
\end{tabular}
\end{table}
\begin{equation}
p_{\text{sep}}=\mathbb{E}_{\boldsymbol{\theta}}\left[\text{sgn}\left(\theta_{k}^{j,i}-\frac{1}{N}\right)\right]\label{eq33},
\end{equation}
where $\boldsymbol{\theta}$ consists of all differences in each AoA pair.

Table~\ref{T1} shows the separation probability as the density of vehicles increases from 10 to 40 cars/km/lane, where $p_{\text{sep}}$ increases highly by the number of antennas. It should be noted that $p_{\text{sep}}$ depends more on the resolution of the array than the vehicle density. Based on Table 1, we would say that the (b) channel scenario happens with about 90$\%$ and (c) happens with about 10$\%$. In channel (b), much of the received energy of two paths are likely to be physically separated at the antenna array by the energy-focusing property of the lens-MIMO. Thus, the inter-path interference prior to SIC is effectively suppressed, which improves the performance. Additionally, we can see that the R2SA could be a feasible solution for the AoA estimation in the considered scenario with several antennas.
\subsubsection{Computational complexity}A key advantage of the R2SA is its lower computational complexity over conventional schemes. Specifically, the total number of multiplication operations for the R2SA with SIC in Algorithm \ref{alg} is as follows: $T_{\text{R2SA}} = K(2N+2)$. This complexity comes from four steps in Algorithm \ref{alg}: the first is to find the antenna index of the strongest received power in $N\times 1$ vector ($N$, line3); the second part is to calculate the ratio of two adjacent antennas and the error (2, line6-7); the third part is to compute the SIC signal with the element-wise product of two $N\times 1$ vectors ($N$, line14); the last part is for the `$for\,loop$' of the SIC iteration ($K$, line1), respectively. 

In contrast, conventional methods like MUltiple SIgnal Classification (MUSIC) and maximum likelihood (ML) require matrix multiplications for the correlation matrix and eigenvalue decomposition. They also involve additional computations for evaluating the pseudo-spectrum and performing exhaustive searching in a preset dictionary \cite{com, est6}. In big-$\mathcal{O}$ sense, MUSIC and ML methods have an approximate number of multiplications of $\mathcal{O}(d_{\text{dic}}N^4)$ and $\mathcal{O}(d_{\text{dic}}N^3)$, respectively, where $d_{\text{dic}}$ represents the size of the pre-defined dictionary. It is readily noted that the complexity of R2SA is significantly lower than conventional methods, and that the limited computing resources of vehicles make the R2SA method more advantageous compared to previous estimation schemes in V2V communications.
\section{AoA-based Localization}\label{sec4-4}
In this section, we first explore the maximum likelihood (ML) localization and propose an alternative method that solves the sensing equations (SEs) of the AoA-based geometric function for the relative localization in the considered intersection scenario. Regarding the position $\mathbf{p}$ and orientation $\boldsymbol{\omega}$ vectors, we focus on the relative localization using the AoA estimates $\hat{\theta}_{k,j}\,\forall k,j\in\mathcal{V}$, where the probability density function (pdf) of the AoA measurement model is known, as depicted in \eqref{eq6}, and the variance of estimated AoA at each vehicle is unknown.

Further, by Bayes' theorem for \eqref{eq6} without the constant terms, which correspond to the prior and marginal probabilities for the AoA and localization parameters, respectively, the conditional joint density function of $\mathbf{p}$ and $\boldsymbol{\omega}$ given by $\hat{\boldsymbol{\theta}}$ can be re-written as
\begin{equation}
    \begin{split}
f&(\mathbf{p},\boldsymbol{\omega}|\hat{\boldsymbol{\theta}})\\
&= \prod_{k\in\mathcal{V}}\prod_{j\in\mathcal{V}_k} f(\mathbf{p}_{k},\mathbf{p}_{j},\omega_{k,j}|\hat{\theta}_{k,j})\\
& = \prod_{k\in\mathcal{V}}\prod_{j\in\mathcal{V}_k}\frac{1}{ \sqrt{2\pi\sigma_{\text{AoA}}^{2}(\hat{\theta}_{k,j})}} \text{exp}{ \left(- \frac{(\alpha_{k,j}-\hat{\theta}_{k,j})^{2}} {2\sigma_{\text{AoA}}^{2}(\hat{\theta}_{k,j})} \right)},\label{eq35}
    \end{split}
\end{equation}
where $\sigma_{\text{AoA}}^{2}(\hat{\theta}_{k,j})$ is CRLB at $\hat{\theta}_{k,j}$ in \eqref{eq4}. Eq.~\eqref{eq35} holds when each target vehicle is assigned to a different sub-channel such that all vehicle communication channels are independent. Following this, we localize the vehicles by maximizing the joint pdf as follows:
\begin{equation}
    [\hat{\mathbf{p}}, \hat{\boldsymbol{\omega}}]=\max_{\mathbf{p}, \boldsymbol{\omega}}\sum_{k\in\mathcal{V}}\sum_{j\in\mathcal{V}_k}\text{log}f(\mathbf{p}_{k},\mathbf{p}_{j},\omega_{k,j}|\hat{\theta}_{k,j}).\label{eq36}
\end{equation}


In \eqref{eq35}, the variance of estimated AoA is often unavailable in estimating localization. Therefore, the other way of localization is to solve a set of SEs, which are defined as $\mathcal{L}_{k,j}=\left| g(\mathbf{p}_k,\mathbf{p}_j,\omega_{k,j})-\hat{\theta}_{k,j}\right|^2\forall k,j\in\mathcal{V}$, assuming all estimates of AoA are accurate (i.e., $\mathcal{L}_{k,j}=0$ for all $k$ and $j$). Consequently, the localization parameters can be obtained as follows:
\begin{equation}
    [\hat{\mathbf{p}},\, \hat{\boldsymbol{\omega}}]=\underset{\mathbf{p},\,\boldsymbol{\omega}}{\min}\sum_{k\in\mathcal{V}}\sum_{j\in\mathcal{V}_k}\left| g(\mathbf{p}_k,\mathbf{p}_j,\omega_{k,j})-\hat{\theta}_{k,j}\right|^2.\label{eq37}
\end{equation}

The solution of \eqref{eq37} approaches ML asymptotically, as SNR increases \cite{ap}. As SNR increases, $\sigma_{\text{AoA}}^{2}$ decreases and each estimate $\hat{\theta}_{k,j}$ gets close to the actual geometric function $g(\mathbf{p}_k,\mathbf{p}_j,\omega_{k,j})$.  Consequently, the SEs $\mathcal{L}_{k,j}\forall k,j$ in \eqref{eq37} tend to approach zero. Thus, all of $\mathcal{L}_{k,j}$ are linearly dependent since the sum of SEs weighted with arbitrary non-zero values $\upsilon_{k,j}$ can be readily shown zero (i.e., $\sum_{k\in\mathcal{V}}\sum_{j\in\mathcal{V}_k}\upsilon_{k,j}\mathcal{L}_{k,j}=0$ for any non-zero $\upsilon_{k,j}$). Hence, the set of sensing equations is not over-determined and possesses a unique solution. Moreover, in the fully-connected vehicular channel case, we have the $3N_v$ unknown parameters ($x_k$, $y_k$, and $\omega_{k,j}\,\forall k$) and the $N_v(N_v-1)$ equations following $\tan\left({\hat{\theta}_{k,j}+\omega_{k,j}}\right)=(y_j-y_k)/(x_j-x_k)\,\forall j,k\in\mathcal{V}$. To obtain a feasible solution of \eqref{eq37} without encountering an under-determined case, we then require $N_v(N_v-1)\geq3N_v$, which leads to at least four vehicles being required, i.e., $N_v\geq4$.
\begin{figure}
 \renewcommand{\figurename}{Fig.}
\centering
\subfloat[Position, $\sum_{k\in\mathcal{V}}\sum_{j\in\mathcal{V}_k}\log\left| g(\mathbf{p}_j|\mathbf{p}_k)-\hat{\theta}_{k,j}\right|^2$]{{\includegraphics[width=0.90\linewidth]{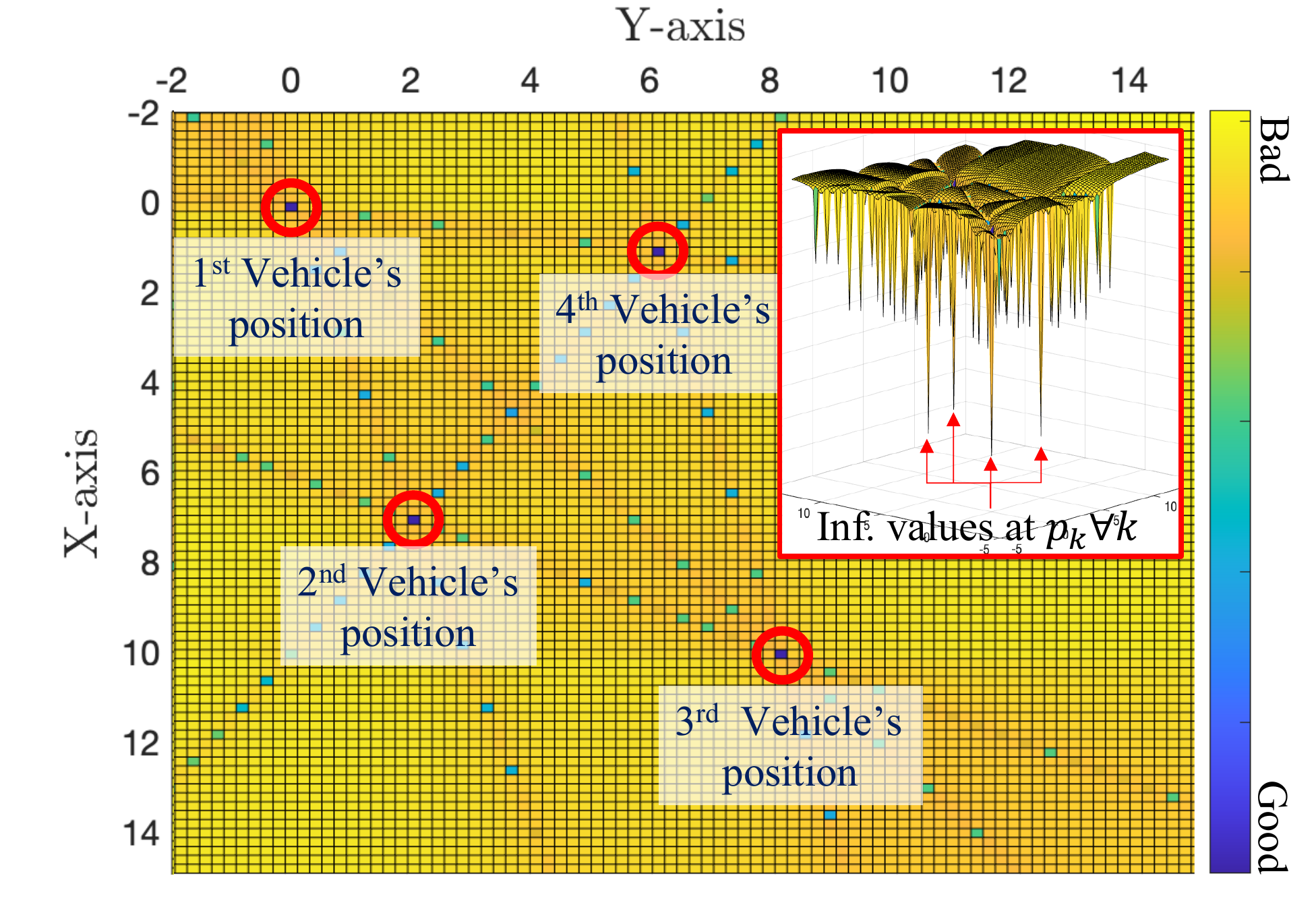} }}

\centering
\subfloat[Orientation, $\left| g(\omega_{k,j}|\mathbf{p}_k)-\hat{\theta}_{k,j}\right|^2$]{{\includegraphics[width=0.95\linewidth]{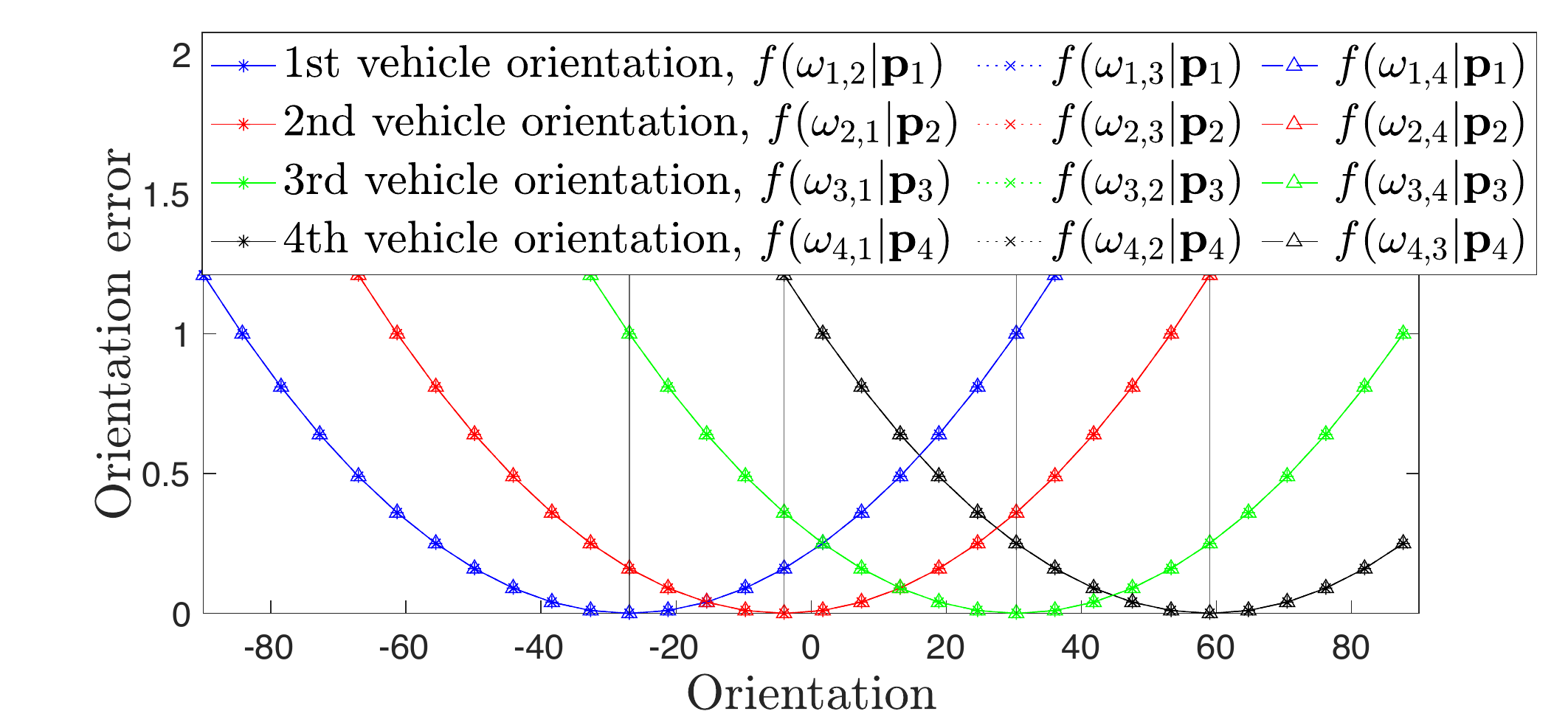} }}
\caption{Sum of sensing equations.}
\label{f6}
\end{figure}

For the relative localization in \eqref{eq37}, the localization parameters may be determined in a centralized manner, where each vehicle estimates the AoAs of its target vehicles and sends them to a mobile edge cloud (MEC) placed in near by the road side unit (RSU). Due to the MEC's ability to process high-volume data in real-time, as indicated in \cite{MEC1,MEC2}, the MEC can estimate the position and orientation of all vehicles with no processing delay. Considering the millisecond-level network latency and the computing power of the MEC, the service latency would be kept below 10 milliseconds, which is significantly smaller than the variations in vehicle's mobility, particularly for short-range V2V scenarios. Each vehicle then has the relative localization knowledge about the target vehicles. If the $k$-th vehicle can acquire its absolute localization in terms of its own sensing capability, then the relative localization information can directly transform into absolute ones. 

Fig. \ref{f6} shows the proposed localization method solving SEs in \eqref{eq37} for the AoA estimates $\hat{\theta}_{k,j}\forall j$ of R2SA when SNR $=5$~dB and $N=61$. To intuitively reveal the solution obtained from \eqref{eq37} in a two-dimensional figure, we further assume that each vehicle is aware of its own location. Thus, the solution of subject vehicle's position and orientation would be equivalent to \eqref{eq37} under the feasible conditions where neither over-determination nor under-determination occurs. Fig. \ref{f6}(a) shows the sum of the logarithm SE for each vehicle's position, where the dotted lines connecting each pair of vehicles represent the SEs $\mathcal{L}_{k,j}$. We can observe the presence of points with infinite negative values at the intersection points of the equations $(g(\mathbf{p}_k,\omega_{k,j}|\mathbf{p}_j)-\hat{\theta}_{k,j})\forall j\in\mathcal{V}_k$. In Fig. \ref{f6}(b), the points with the minimum value of the squared error exist at the real orientation of each vehicle, where those values for each connected vehicle are all the same due to ignoring AoA variance. It is noted that solution of \eqref{eq37} is equivalent to that of the ML localization, and is being adopted as a feasible solution for cooperative localization without over-and-under-determined problems.

It is noted that the position and orientation can be jointly estimated by solving the sensing equations in \eqref{eq37}, without relying on channel quality information and other measurements such as the time of arrival (ToA) and phase difference of arrival (PDoA).
\begin{remark}
(Absence of line-of-sight (LoS) path)
In the scenario of an intersection street, short-distance V2V communications often benefit from the availability of a LoS path \cite{c14-1, losp1, losp2}. However, it is possible for the LoS path to be obstructed by nearby vehicles. There have been papers addressing the challenge of localizing hidden vehicles, such as \cite{b15}, where the positions of scatters and hidden vehicles are jointly estimated under the assumption of sufficiently strong multipath signals. While this approach can be adapted for lens-MIMO systems, it does come with the drawback of increased computational complexity for vehicles.

Hence, we can explore alternative approaches that are more suitable and applicable in the intersection scenario, offering potential improvements over existing methods. In situations where the LoS channels of certain target vehicles are obstructed by nearby vehicles in close proximity, the signal power received from those channels tends to experience significant attenuation due to scattering and absorption losses. This attenuation of received signal power provides a valuable indicator for determining whether the V2V link is blocked or not. As a result, we can exclude the sensing equations that involve channels with low signal power in equation \eqref{eq37}. As long as the number of remaining sensing equations is larger than the number of unknown parameters (i.e., $N_v(N_v-1) - N_{\text{disc}}\geq3N_v$, where $N_{\text{disc}}$ represents the discarded sensing equations), the localization method in \eqref{eq37} can still find a feasible solution. The specific scheme for eliminating the LoS channel is an interesting direction for future research.
\end{remark}

\section{Numerical results}\label{sec5}
In this section, we evaluate the performance of the proposed localization algorithm in the street intersection scenario, where single or multiple target vehicles are allocated in the same sub-channel. Their results are compared with the CRLBs and also with target requirements for 5G positioning service in 3GPP, where position and orientation accuracy should be less than 0.2 m and $2^{\circ}$ with 95 $\%$ confidence level \cite{req}. In order to make the confidence analysis easier, we further assume that the estimate of position and orientation are Gaussian random variables, where the 95$\%$ confidence means the estimate error is within $\pm1.96\sigma$ when estimation error variance is $\sigma$. We also evaluate the performance impact of the number of vehicles assigned in the same sub-channel, and system conditions to satisfy the requirements for the 5G positioning services.

\subsection{Parameter setup}
Fig. \ref{f1}(a) depicts the street intersection with each lane width of 5 m and length of 30 m, where there are three roads, and each road has two lanes per direction. We adopt a communication radius of $R=50$~m, as specified by the 3GPP standardization for the urban intersection scenario \cite{50m}. Specifically, the vehicles are randomly distributed along each lane by a Poisson point process (PPP) with a vehicle density of 10~cars/km/lane, where the length and width of vehicles are 4.7 m and 1.8 m, respectively, and the PPPs for the different lanes are assumed to be independent. The carrier frequency is 28~GHz and the complex, and the complex channel gain $h_{k,j}$ in \eqref{eq1} follows a distribution of $\mathcal{CN}(0,1)$ \cite{vh,cha}. Additionally, the AoA $\theta_{k,j}$ and relative orientation $\omega_{k,j}$ are determined by the geometry of the placement of vehicles. The path loss model of each LoS link, denoted by $\rho_{\text{o}}$, is assumed as follows by geometric statistics \cite{b10}, it is given by
\begin{equation}
    \frac{1}{\rho_{\text{o}}^{k,j}}=\zeta^{2}(d_{k,j})\left(\frac{\lambda}{4\pi d_{k,j}}\right)^2,\label{eq38}
\end{equation}
where $\zeta^{2}(d_{k,j})$ is the atmospheric attenuation over distance $d_{k,j}$. Lens-MIMO and ULA have $(2L+1)$ antenna elements, whose spacing is $f/L$ and $\lambda/2$, respectively. In order to provide complete angular coverage for the vehicle, we assume that two lens-MIMO systems are mounted on the front and back of the vehicle, each capable of an angular view from $-90^{\circ}$ to $90^{\circ}$.

\subsection{Simulation results}
We evaluate the proposed AoA-based localization algorithm in two cases: (a) the single target AoA estimation in Section \ref{sec4-2} and (b) the multi-target AoAs estimation in Section \ref{sec4-3}. In the case of (a), we confirm that the proposed algorithm approaches the CRLBs of the position and orientation, and compare it with the target requirements for 5G positioning services. In the case of (b), we analyze the feasibility of lens-MIMO-based localization for multi-target AoAs estimation in the V2V street intersection scenario.

\subsubsection{A single target in a sub-channel}\label{single}We consider the unicast of each vehicles in different sub-channels. To evaluate the performance of the proposed scheme, we adopt the root mean squared error (RMSE) of the  position and orientation estimates, denoted as $\text{RMSE}_{p}$ and $\text{RMSE}_{\omega}$, respectively. These metrics are as follows:
\begin{equation}
    \text{RMSE}_{p}=\frac{1}{N_{V}}\sum_{k\in\mathcal{V}}\sqrt{||\mathbf{p}_k-\hat{\mathbf{p}}_k||^2},\label{eq39}
\end{equation}
\begin{equation}
    \text{RMSE}_{\omega}=\frac{1}{N_{V}}\sum_{k\in\mathcal{V}}\sqrt{|\omega_{k,j}-\hat{\omega}_{k,j}|^2}.\label{eq40}
\end{equation}
\begin{figure}[!t]
 \renewcommand{\figurename}{Fig.}
\centering
\subfloat[Position]{\includegraphics[width=0.45\textwidth]{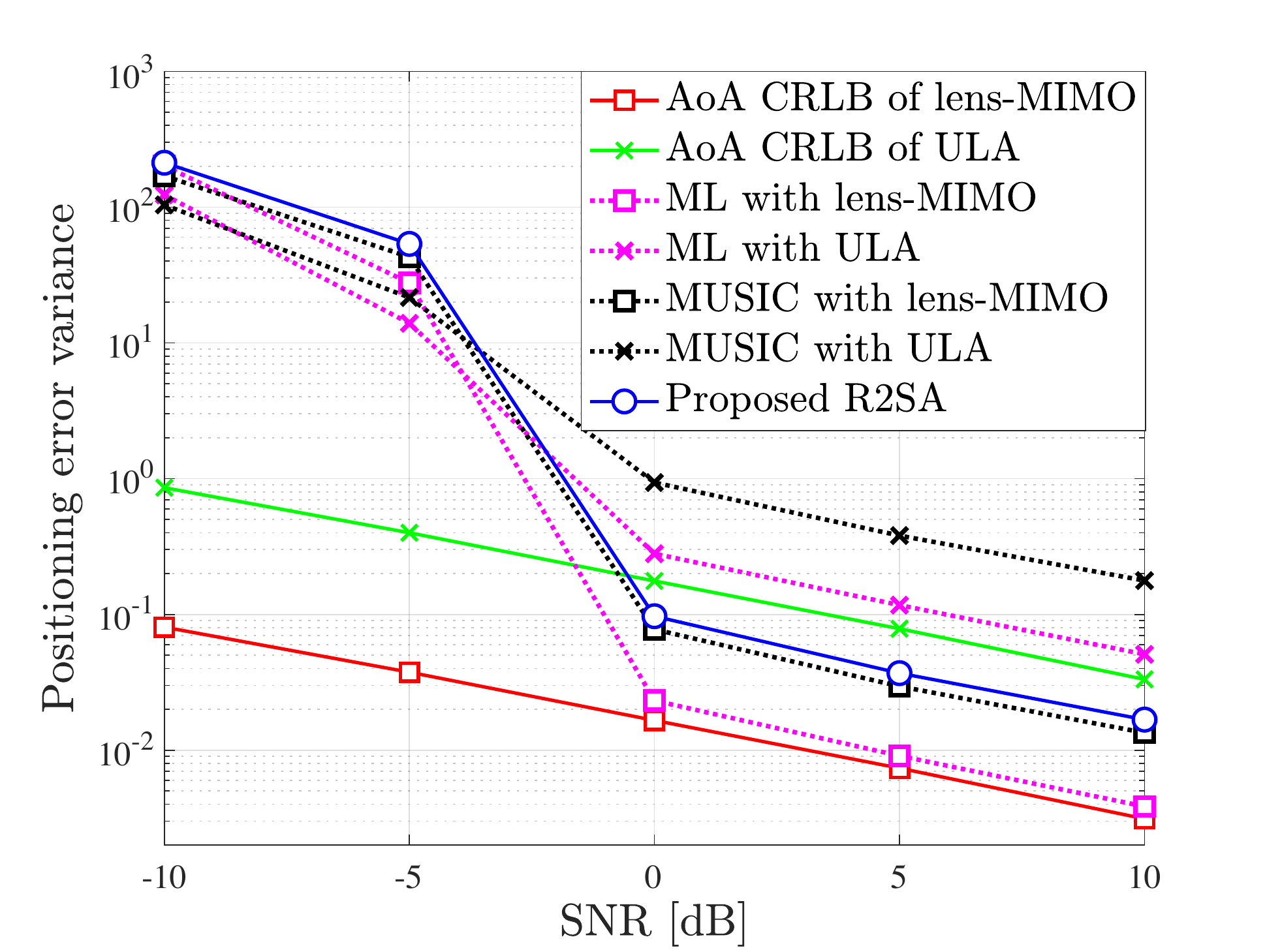}}\\
\centering
\subfloat[Orientation]{\includegraphics[width=0.45\textwidth]{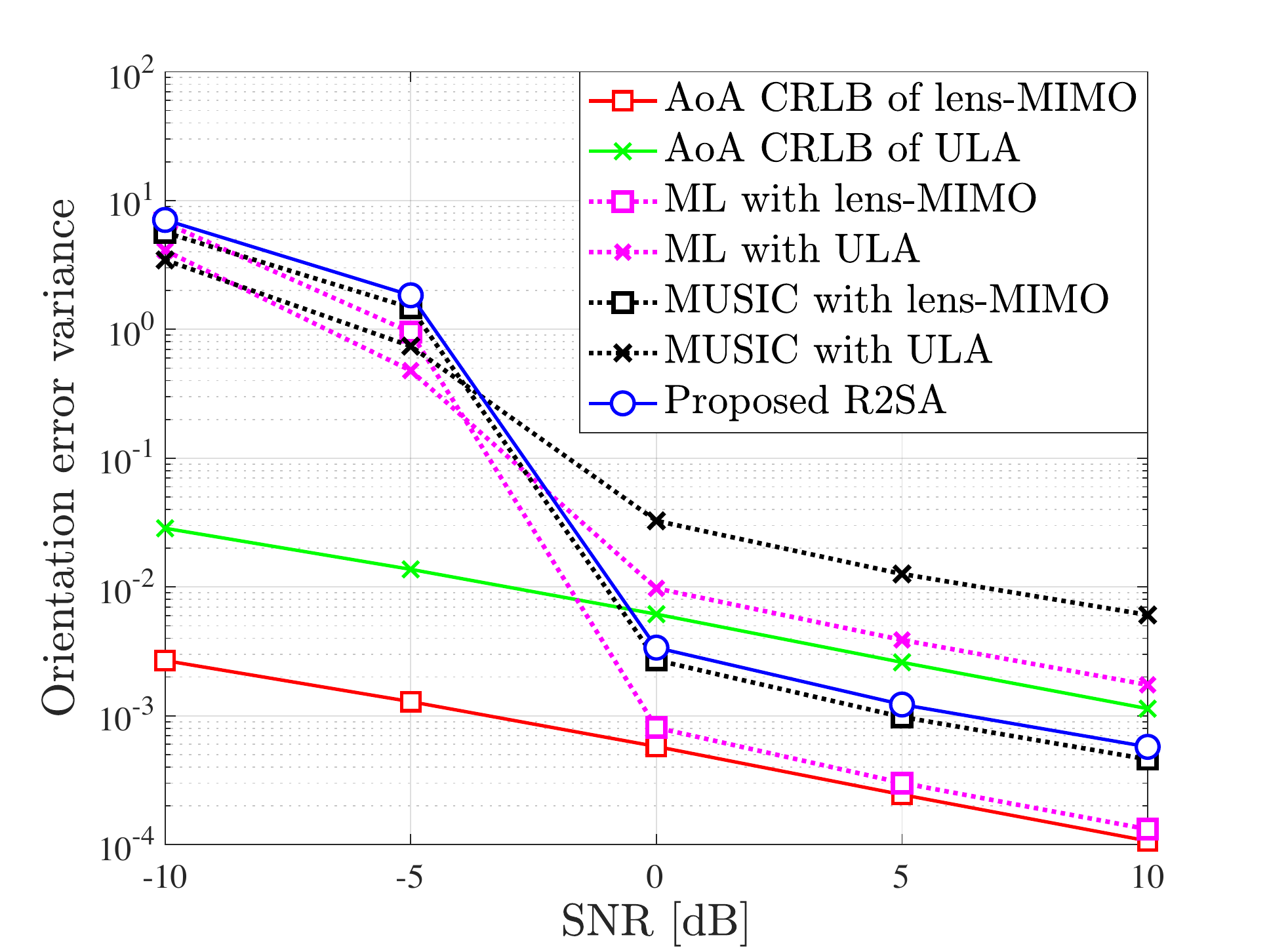}}
\caption{Comparison of localization errors among different methods for AoA estimation.}
\label{estcom}
\end{figure}
Fig. \ref{estcom} presents the localization error variance as the SNR increases, and compares the proposed R2SA with the existing methods, such as MUSIC and ML estimations involving exhaustive searching within a dictionary range of $[-90^{\circ}, 90^{\circ}]$ with a resolution of $0.1^{\circ}$, where positions and orientations are estimated by \eqref{eq37}. The number of antennas and vehicles are set up to 121 and 4, respectively. The focal length is set to $30\lambda$, which is the minimum focal length for the lens aperture $L=60\lambda$ to satisfy the condition $f\geq L/2$. The AoA CRLBs of lens-MIMO and ULA in \eqref{eq4} and \eqref{eq3} are depicted as dashed red and greenish lines, respectively. The MUSIC method is represented by dotted black lines for lens-MIMO and ULA, while the ML method is represented by dotted magenta lines. The proposed R2SA in \eqref{eq31} is indicated by the dashed blue line.

Fig. \ref{estcom} also verifies that ML method for both position and orientation approaches the derived CRLBs of position and orientation in \eqref{eq17} and \eqref{eq18}, which provides the authenticity of the derived CRLBs. In the low SNR region ($\text{SNR}\leq-4$~dB), the performance of all estimators utilizing lens-MIMO is adversely affected since its received peak power feature at the array is no longer visible in the high power noise. However, as the SNR increases, lens-MIMO surpasses the ULA in performance for both the MUSIC and ML methods. For SNR above $-1$~dB, the proposed R2SA algorithm exhibits superior performance compared to the CRLB of ULA, and its performance gets close to the MUSIC method. From the computational complexity perspective discussed in Section \ref{per}, R2SA demonstrates significantly lower computational complexity, approximately $2\times10^{2}$ ($\mathcal{O}(2N)$) per target vehicle. In contrast, the ML and MUSIC methods require approximately $4\times10^{11}$ ($\mathcal{O}(d_{\text{dic}}N^4)$) and $3\times10^{9}$ ($\mathcal{O}d_{\text{dic}}N^3$) multiplications, respectively, for each vehicle. Hence, R2SA can be considered more suitable than other methods for V2V systems with limited computing resources, as it offers real-time signal processing and low latency.
\begin{figure}[!t]
 \renewcommand{\figurename}{Fig.}
\centering
\subfloat[Position]{\includegraphics[width=0.46\textwidth]{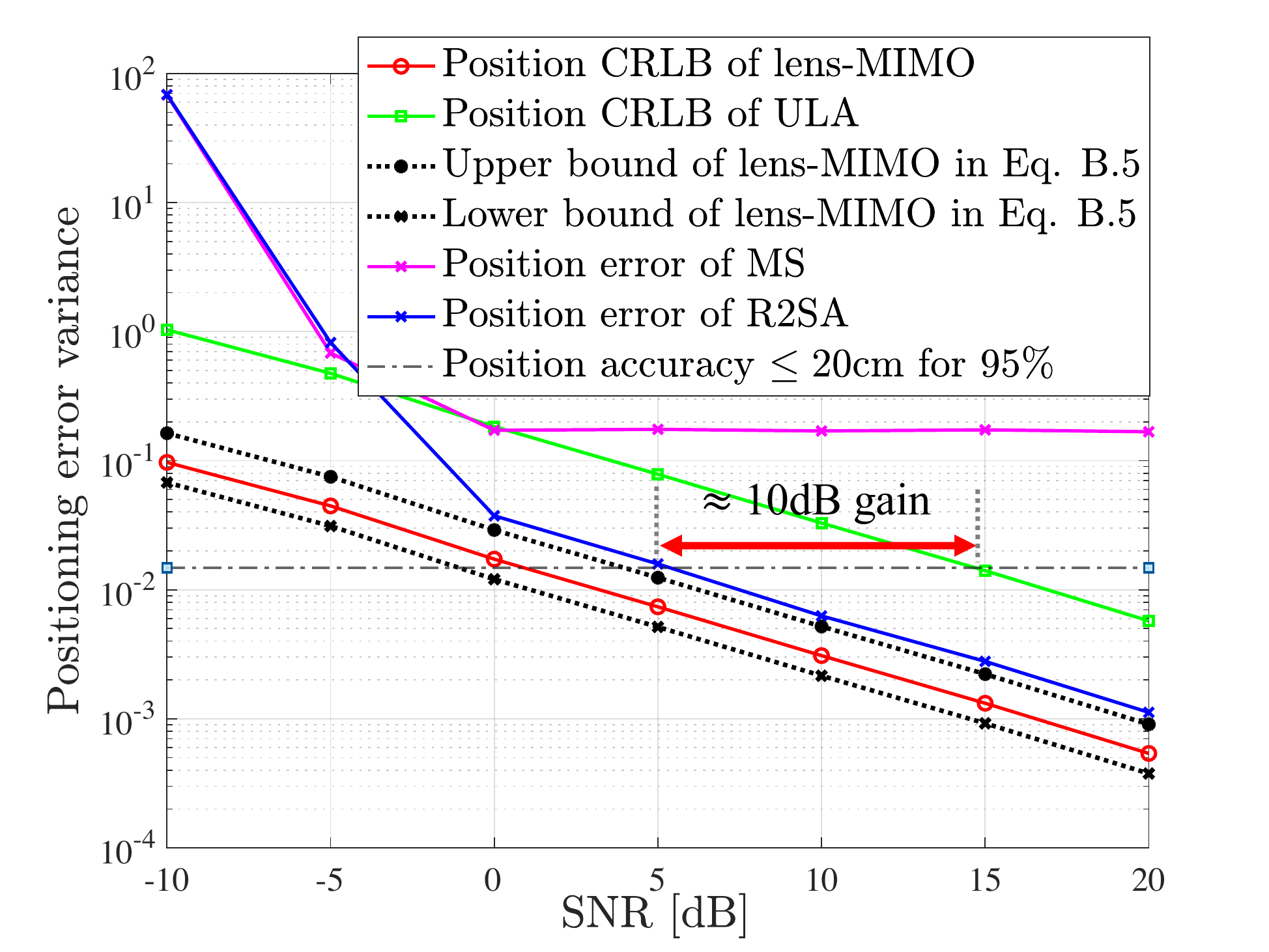}}\\
\centering
\subfloat[Orientation]{\includegraphics[width=0.46\textwidth]{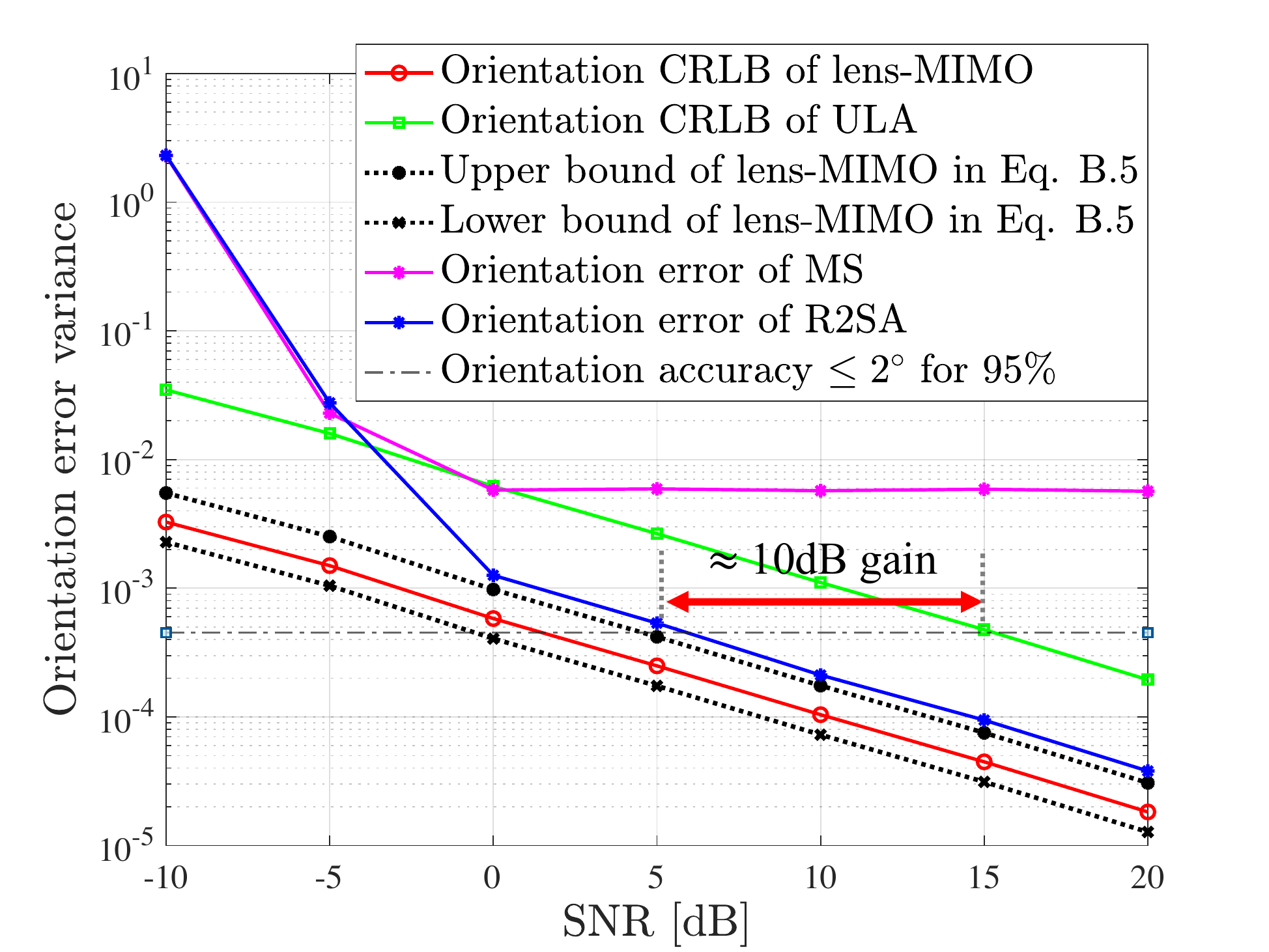}}
\caption{Comparison of localization errors among different derived CRLBs and target requirements.}
\label{f7}
\end{figure}

In Fig. \ref{f7}, we compare the CRLBs of the localization parameters (position $\mathbf{p}$ and orientation $\boldsymbol{\omega}$) for lens-MIMO and ULA as the SNR increase, with the same simulation parameters of Fig. \ref{estcom}. 
The dashed red and dotted greenish lines are the CRLB of lens-MIMO and ULA, respectively. The dotted black lines are the derived upper and lower bound, and dotted blue and flushed lines are the localization performance with AoA estimates of R2S2 in \eqref{eq31} and MS in \eqref{eq30}, respectively. We also compare these performances with the positioning requirements of $95\%$ confidence in 0.2~m position and $2^{\circ}$ orientation accuracy, which are shown in horizontal black dash-dotted lines.

Fig. \ref{f7} verifies that the position and orientation CRLB of lens-MIMO in \eqref{eq17} and \eqref{eq18} are always superior to the CRLB of the conventional ULA by about 13~dB, as shown in \eqref{eq24}. The performances of both position and orientation with R2SA get close to the upper bound of lens-MIMO's CRLB within a marginal gap of about 1.5~dB as the SNR increases. R2SA also satisfies the target requirements of position and orientation at 5 dB, and its performance is about 10~dB better than the ULA's CRLB, even with lower complexity. Meanwhile, the CRLBs of ULA satisfy the requirements at 15~dB. The MS scheme estimates the AoA by quantizing the incoming directions represented by each antenna element. As a result, it acts as a spatial quantizer for the AoA. Consequently, the MS scheme exhibits an error floor at high SNR regions due to the spatial quantization error. It can be noted that the proposed R2SA holds feasibility for 5G positioning services, particularly in the SNR region higher than 5~dB, which can be reliably achieved in mmWave-based V2V communication systems \cite{c23, c14-1}.
\begin{figure}[!t]
 \renewcommand{\figurename}{Fig.}
\centering
\subfloat[Position]{\includegraphics[width=0.46\textwidth]{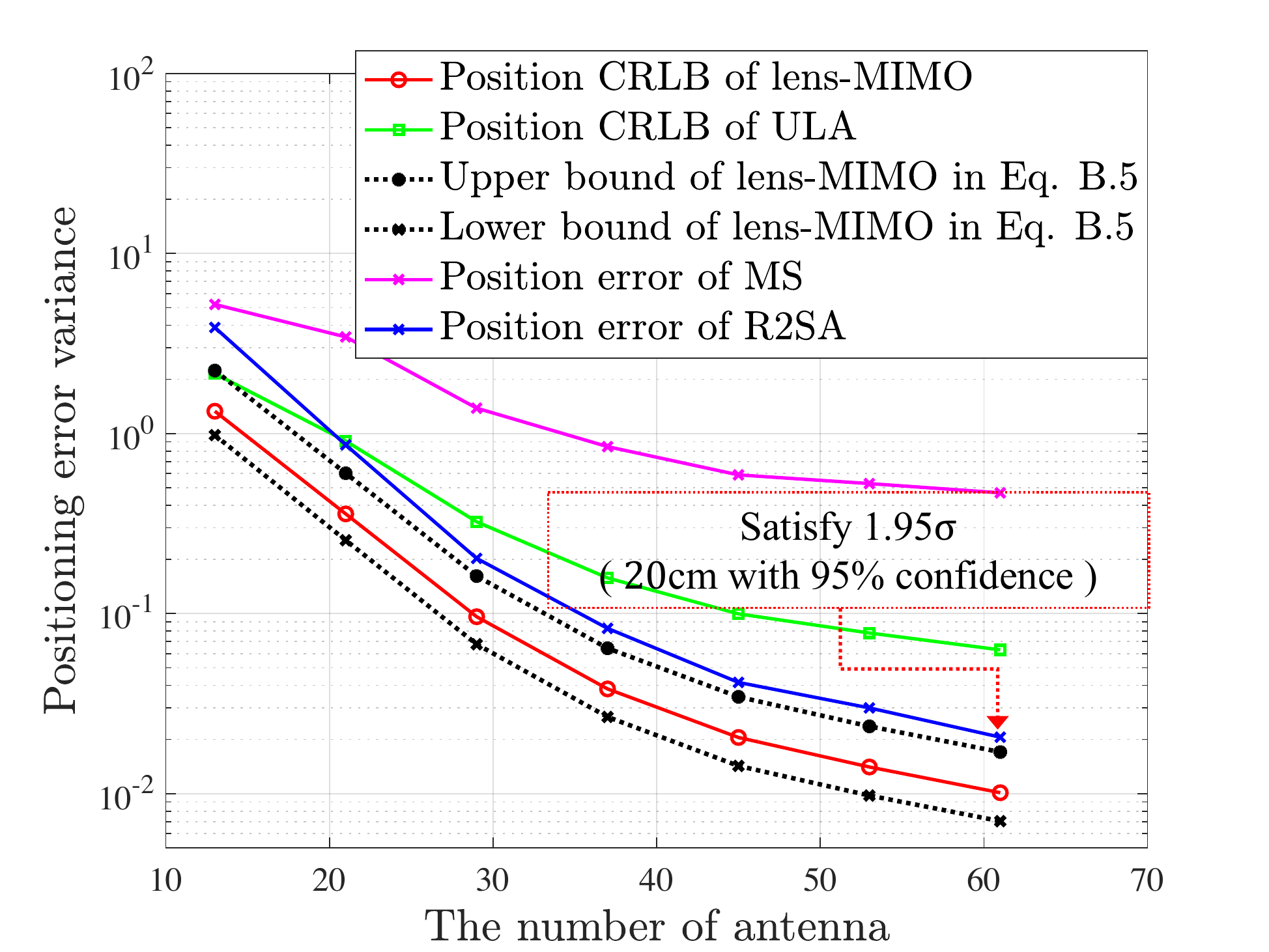}}\\
\centering
\subfloat[Orientation]{\includegraphics[width=0.46\textwidth]{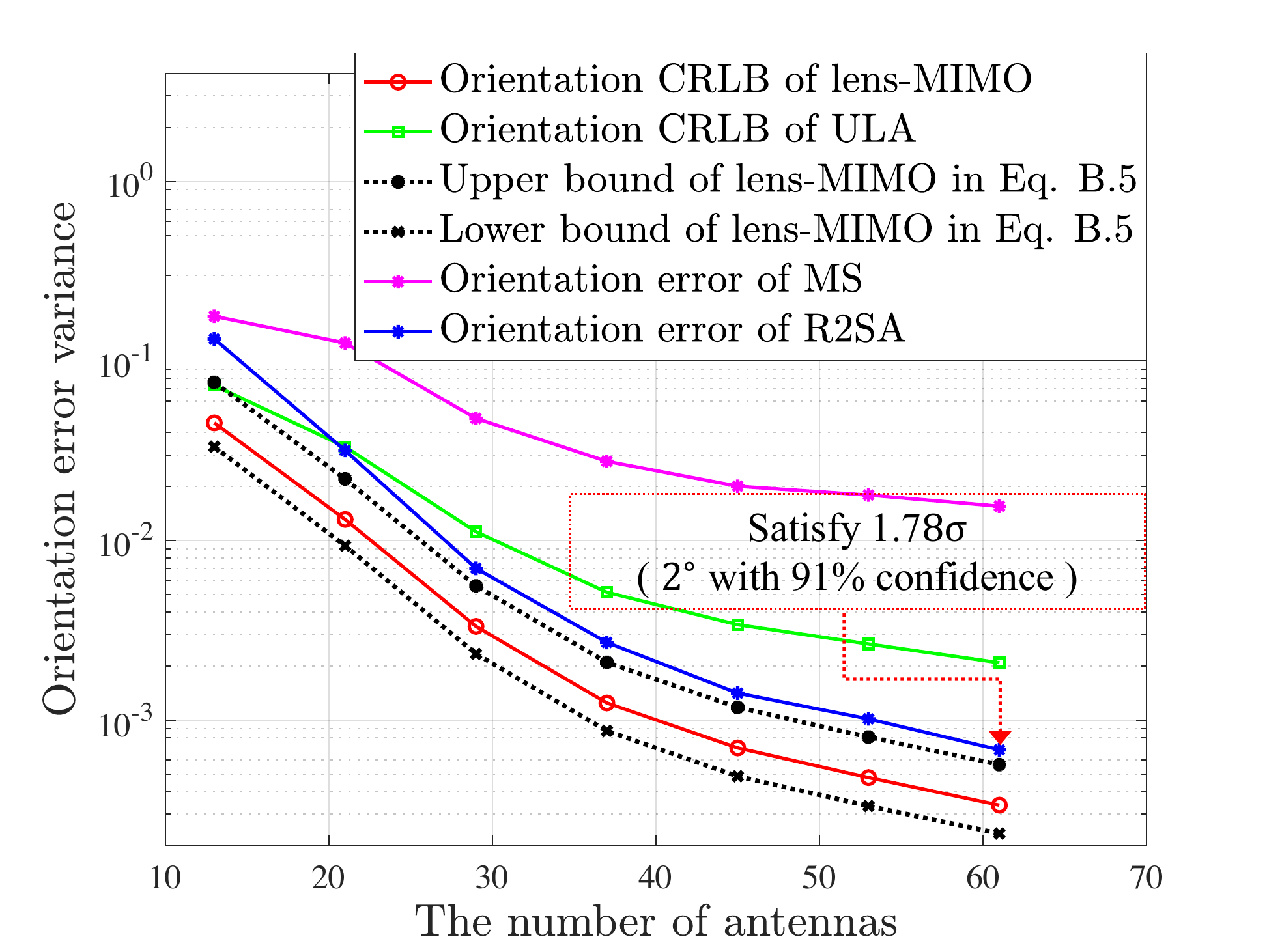}}
\caption{Localization error as the number of antennas increases.}
\label{f8}
\end{figure}

Fig. \ref{f8} compares the same performance curves with Fig. \ref{f7} as the number of antennas increases, where SNR is 10~dB and focal length is the same. For a mmWave lens-MIMO, as the number of antennas would be normally large. Therefore, we consider up to 60 antennas, which can be fit into normal size vehicles, as demonstrated in previous validation studies \cite{cb1, imp}. Particularly, the CRLBs of lens-MIMO and ULA are close for the smaller number of antennas, such as $N=13$, but gets far different as its number of antennas increases as shown in Fig. \ref{f8}(a) and (b). This is because the larger-sized lens-MIMO can receive more energy and achieve higher spatial sampling resolution. In contrast, when lens-MIMO has antenna elements of less than 13, the spatial energy could not be sufficiently collected for all directions since the lens-MIMO can collect only as many spatial samples as the number of antennas.

With more than 20 antennas, the position and orientation accuracy of R2SA is better than CRLBs of ULA, but cannot converge to lens-MIMO's CRLB. Its loss is first due to the fact that R2SA uses only the ratio of two adjacent antennas. Secondly, it is because the sensing equation ignores the measurement noise variance. Notwithstanding, the position accuracy with R2SA satisfies the target accuracy requirement, and its orientation estimate achieves more than 90 $\%$ confidence level at $N=61$, which can be further enhanced by larger number of vehicles involved in the cooperative localization. It is important to note that the proposed R2SA method approaches the target requirements for position and orientation estimation while maintaining a lower complexity compared to utilizing the entire antenna array.
\begin{figure}[!t]
 \renewcommand{\figurename}{Fig.}
\centering
\subfloat[Position]{\includegraphics[width=0.46\textwidth]{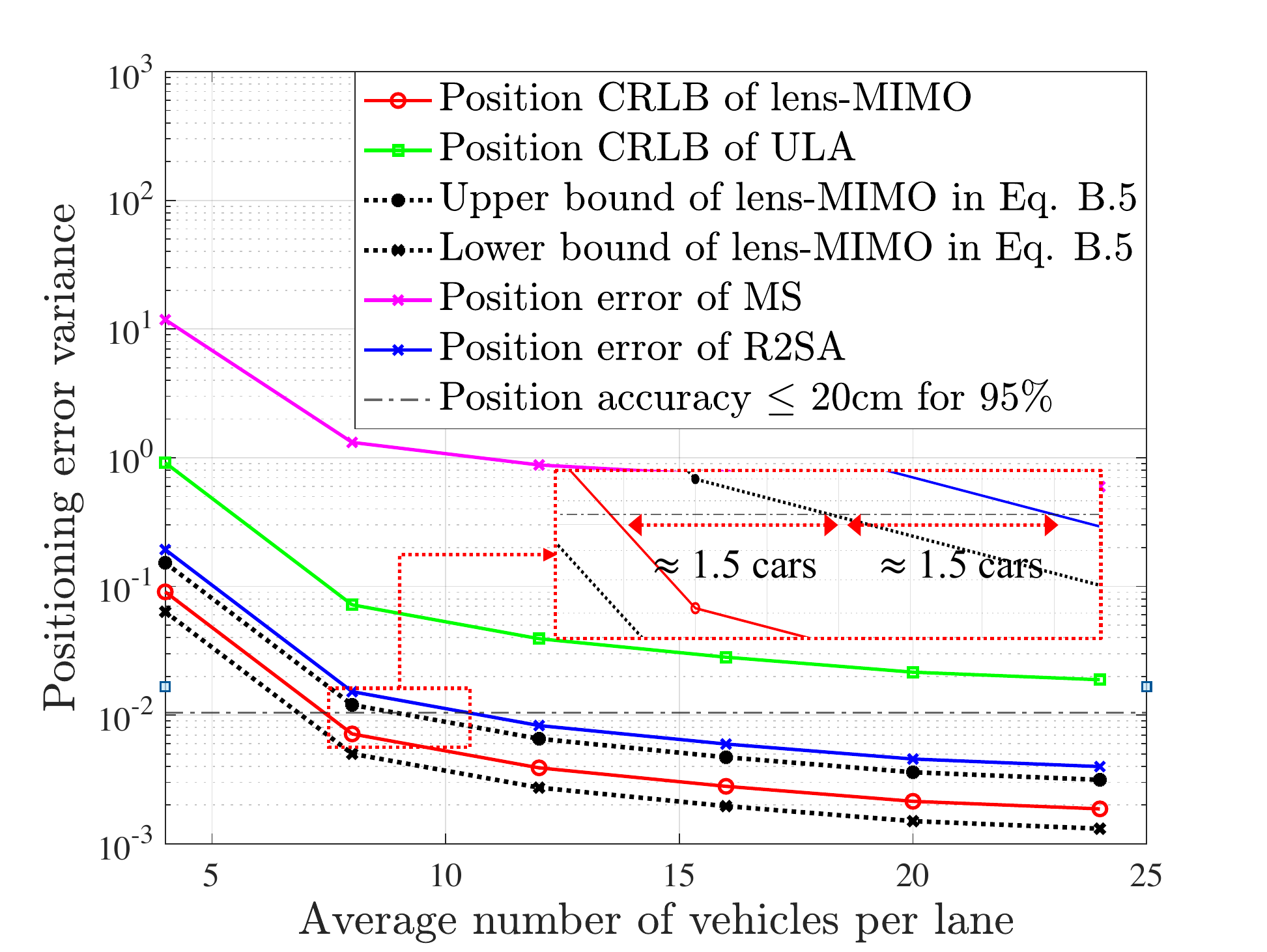}}\\
\centering
\subfloat[Orientation]{\includegraphics[width=0.46\textwidth]{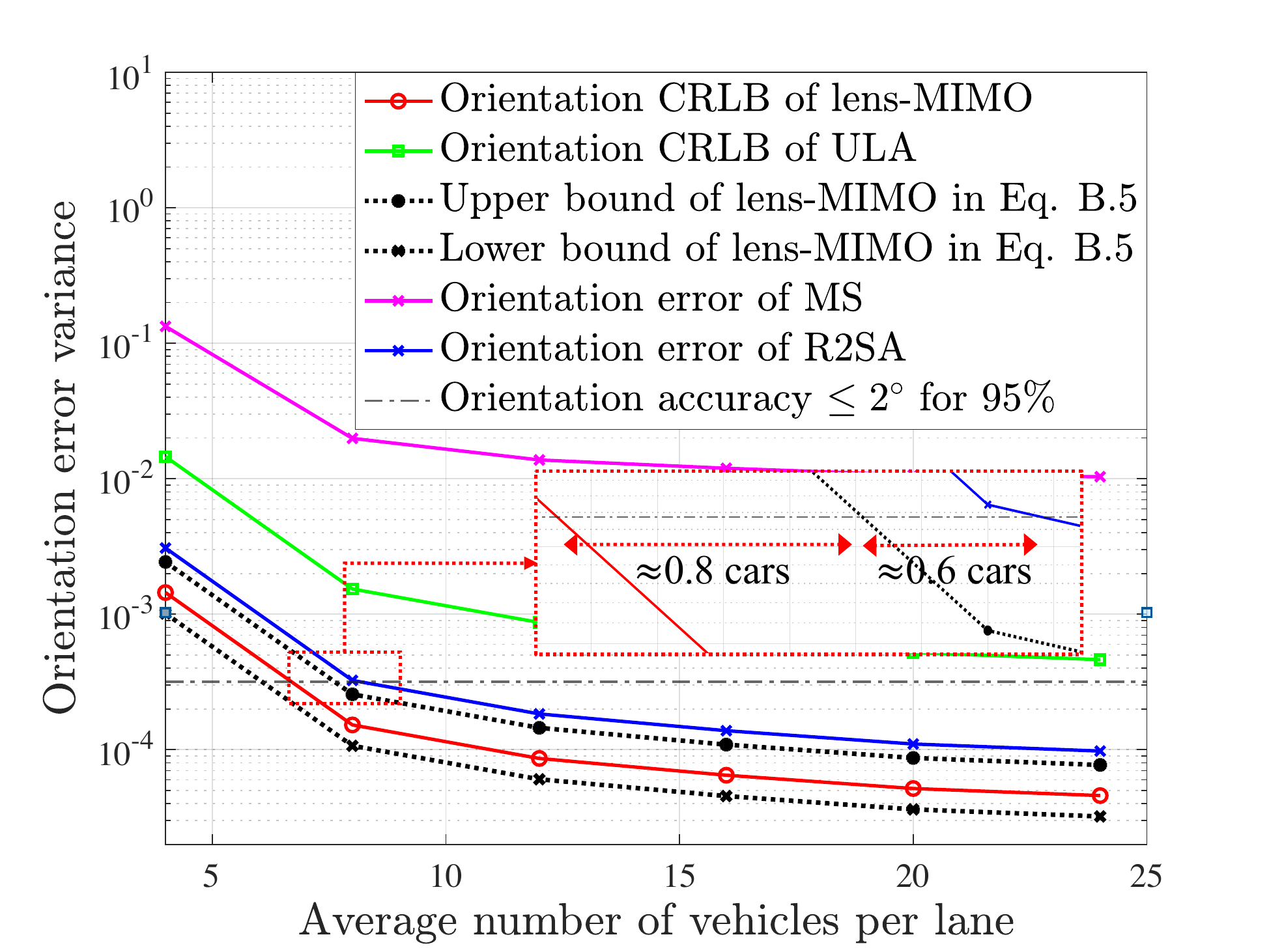}}
\caption{Localization error as the number of vehicles increases.}
\label{f9}
\end{figure}

Fig. \ref{f9} compares the same set of performance curves, which are identical to those shown in Fig. \ref{f8}, for larger number of vehicles participating in localization estimation, where $N$ is 31 and SNR is 10~dB. To increase the total number of vehicle pairs, we raise the average number of vehicles, which is the mean of PPP, resulting in a higher vehicle density. Each vehicle is then sequentially connected to its adjacent vehicles. The results confirm that the position and orientation performance is proportional to the number of vehicles. The R2SA satisfies the target requirements of position and orientation  for vehicle densities of 10 and 8, respectively, whereas CRLBs of ULA fail to meet the demands, even if many vehicles participate in cooperative localization. This shows that R2SA can satisfy the requirements, even with a smaller number of vehicles. Fig. \ref{f9}(a) and (b) verify that the CRLBs of lens-MIMO is in-between the derived upper and lower bound in \eqref{b5}, whose differences in the number of required vehicles for target demands between the CRLBs and upper bounds are insignificant with 1.5 vehicles for position and 0.8 vehicles for orientation. The R2SA requires 3 and 2 more vehicles for the position and orientation accuracy than the CRLB of lens-MIMO.

In the single target case, Figs. \ref{f7}-\ref{f9} remark that the accuracy of the AoA estimation is critical to the localization performance. From the perspective of the target requirements for 5G positioning service, the R2SA of position accuracy is robust for fewer antenna elements in Fig. \ref{f8}, and the R2SA of orientation is suitable with sparser connected vehicles in Fig.~\ref{f9}.

\subsubsection{Multiple vehicles connection in a sub-channel}\label{multi}Next, we consider the multi-target vehicles' AoA estimation in Fig. 1(a), where the vehicles in the street intersections are multi-casting each other in the same sub-channel simultaneously. More than two multiple vehicles are sequentially allocated to the sub-channel until the AoA of all surrounding vehicles is estimated. The vehicle density on each lane is assumed to be 10 cars/km/lane with PPP. In multiple vehicles localization, we explore the RMSE localization error as 
\begin{equation}
\text{RMSE}_{l}=\frac{1}{N_v}\sum_{k\in\mathcal{V}}\sqrt{||\boldsymbol{\eta}_k-\hat{\boldsymbol{\eta}}_k||^{2}},\label{indl}
\end{equation}
where $\boldsymbol{\eta}_k=[x_k, y_k, \omega_{k,j}]$.
\begin{figure}[!t]
 \renewcommand{\figurename}{Fig.}
\centering
{\includegraphics[width=0.97\linewidth]{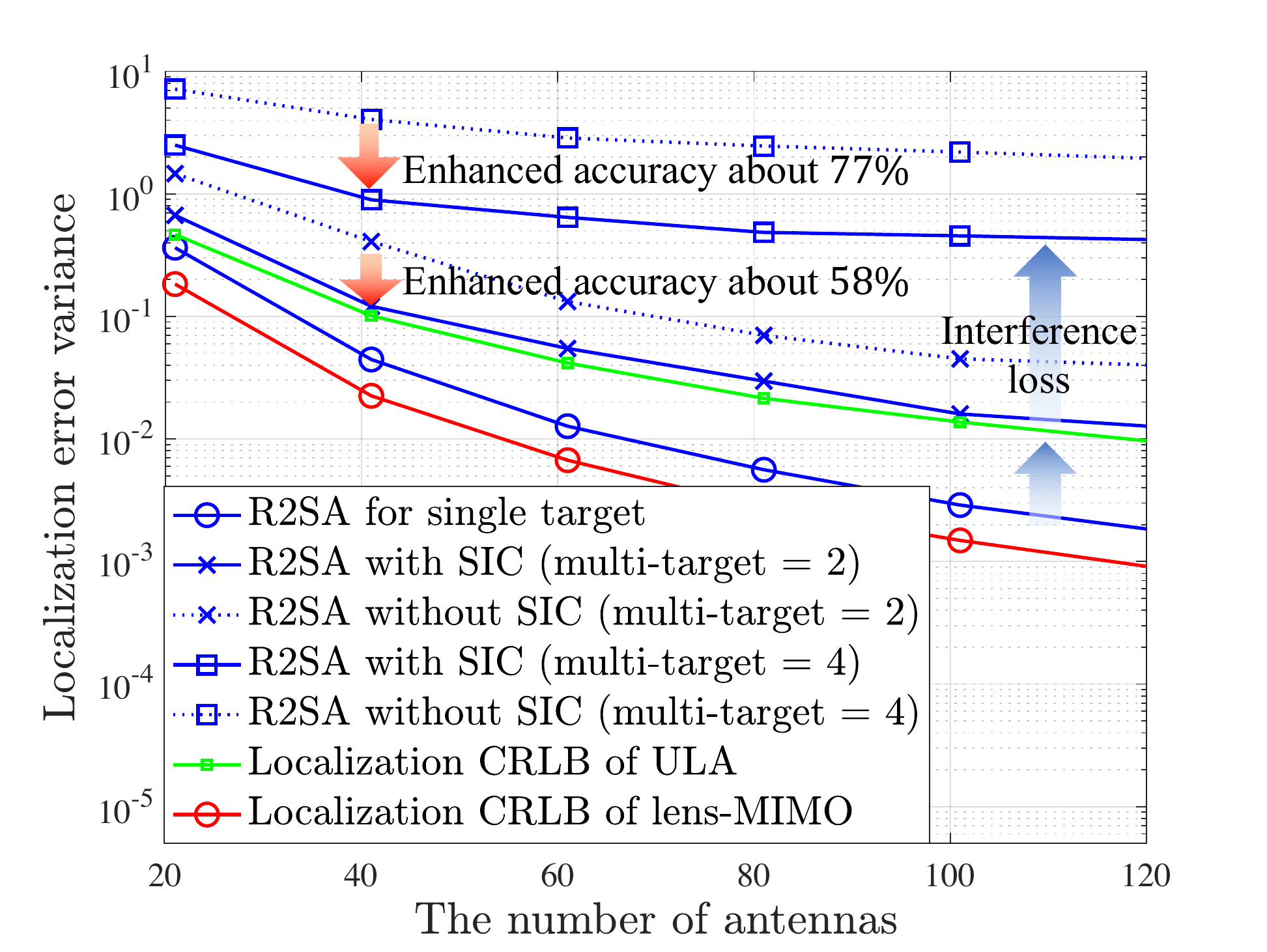}}
\caption{Localization error versus the number of antennas for multi-target $\in\left\{1,2,4\right\}$ vehicles in one sub-channel.}
\label{f10}
\end{figure}

Fig. \ref{f10} shows the localization error variance of multiple AoAs estimation, where eight vehicles are participating in the localization and SNR is 10~dB. The dashed and dotted bluish lines show the localization error variance for R2SA with and without SIC, where we consider the cases that the number of connected target vehicles among eight vehicles in the same sub-channel is one, two, and four. For any number of antennas, the localization error variances with SIC are reduce to approximately 58~\% and 77~\% of ones without SIC for two and four target vehicles, respectively. For two target vehicles, it is worth noting that the multi-AoA estimation for R2SA with SIC approaches to the CRLB of a single AoA for ULA.

However, it is shown that when the number of connected target vehicles is four and $\text{SNR}=10~\text{dB}$, the proposed localization for the R2SA with SIC could not meet the required mean squared error (i.e., $10^{-2}$ for position and $3\times10^{-4}$ for orientation) even for very large number of antennas. Considering the mean squared error variance requirements, we statistically evaluate the feasibility of R2SA with SIC in terms of outage probability that either position or orientation requirement is not supported for a given target. It is defined as
\begin{equation}
\begin{split}
p_{\text{out}}=\mathbb{E}_{\boldsymbol{\eta}}\bigg[\text{sgn}&\left(\sqrt{||[\boldsymbol{\eta}_k]_{1:2}-[\hat{\boldsymbol{\eta}}_k]_{1:2}||^{2}}-\gamma_{p}\right)\\
&\oplus\text{sgn}\left(\sqrt{|[\boldsymbol{\eta}_k]_3-[\hat{\boldsymbol{\eta}}_k]_3|^{2}}-\gamma_{\omega}\right)\bigg],
\end{split}    
\end{equation}
where $A\oplus B$ is a boolean addition of $A$ and $B$, which is zero only when both $A$ and $B$ are zero; $\gamma_{p}$ and $\gamma_{\omega}$ are the target threshold of position and orientation, respectively, determined by $1.96\sigma_p$ and $1.96\sigma_{\omega}$ with accuracy $\sigma_p=20$ cm and $\sigma_o=2^{\circ}$. The sigmoid function $\text{sgn}(x)$ has 1 when $x\geq0$ and 0 otherwise.
\begin{figure}[!t]
 \renewcommand{\figurename}{Fig.}
\centering
\subfloat[$N=121$]{\includegraphics[width=0.99\linewidth]{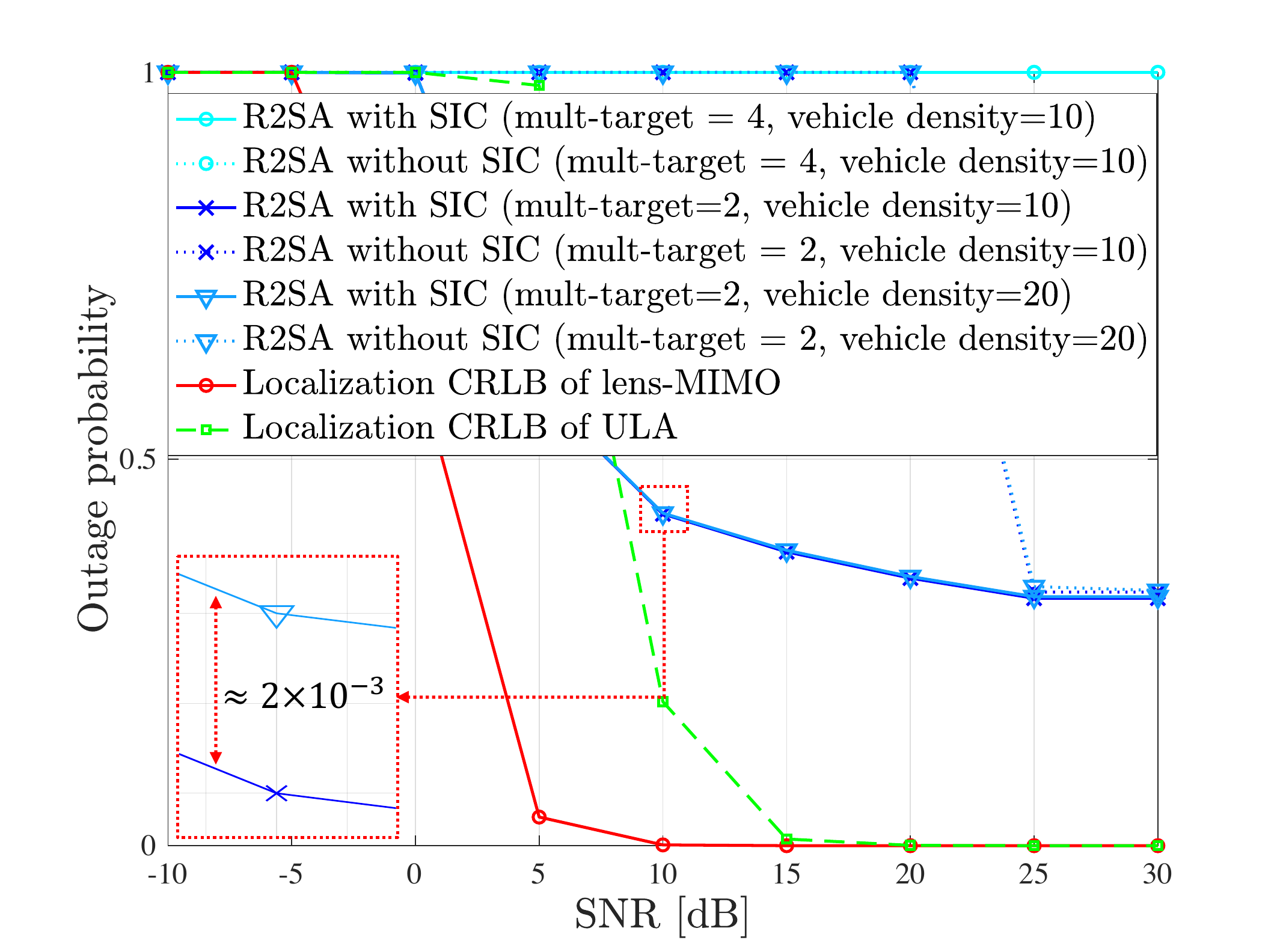}}\\
\centering
\subfloat[$N=161$]{\includegraphics[width=0.99\linewidth]{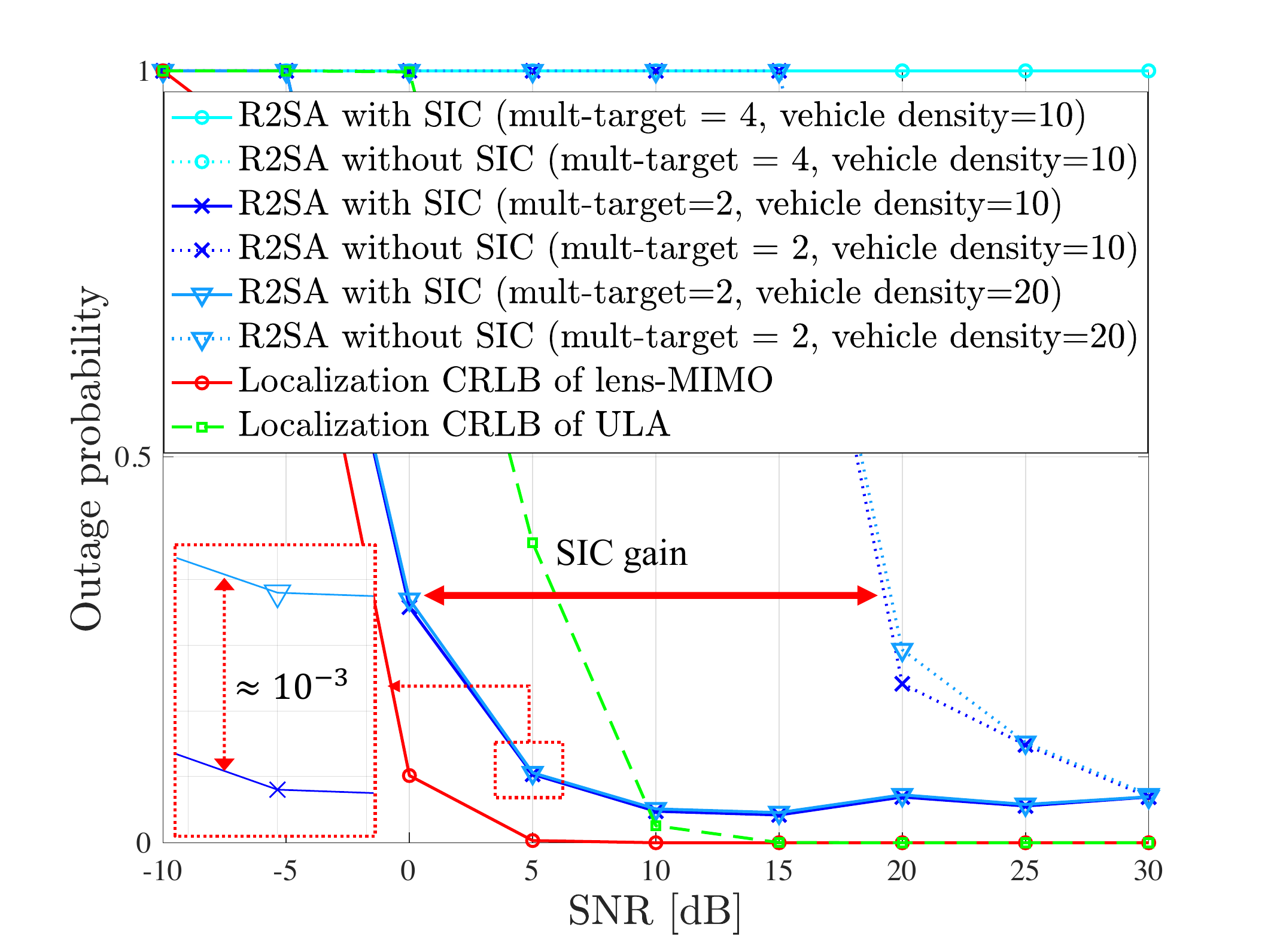}}
\caption{Outage probability versus SNR for multi-target $\in\left\{1,2,4\right\}$ vehicles and vehicle density $\in\left\{10,20\right\}$ cars/km/lane with massive lens-MIMO.}
\label{fadd}
\end{figure}

To satisfy the mean squared error variance requirements, Fig. \ref{fadd} investigates the proposed localization method with R2SA in massive lens-MIMO systems with $N=121$ and $N=161$. We consider different numbers of target vehicles in the same sub-channel and varying vehicle density per lane, while keeping the parameters of the intersection scenario consistent. Fig. \ref{fadd} shows that R2SA with and without SIC ends up with the same outage probability for high SNR region due to the ability of the massive lens-MIMO to effectively suppress inter-path interference through energy focusing, as discussed in Section \ref{per}. We also observe that the performance difference between the vehicle density of 10 and 20 cars/km/lane is negligible. The R2SA with SIC for two target vehicles shows the outage probability floor in Fig. \ref{fadd}(a) for $N=121$. Meanwhile, it achieves the 5 $\%$ outage probability of 20 cm and $2^{\circ}$ at $\text{SNR}=10~\text{dB}$ in Fig. \ref{fadd}(b) for ultra-dense antennas $N=161$, while the SIC gain has about 20~dB for the performance. On the other hand, for the R2SA of four target vehicles in the same sub-channel, it is difficult to keep the 95~$\%$ confidence in the street intersection scenario. We remark that localization with R2SA is affected by the number of vehicles allocated to one sub-channel and the number of antennas rather than the density of vehicles.
\begin{figure}[!t]
 \renewcommand{\figurename}{Fig.}
\centering
\subfloat[Less antenna ($N=31$)]{\includegraphics[width=0.97\linewidth]{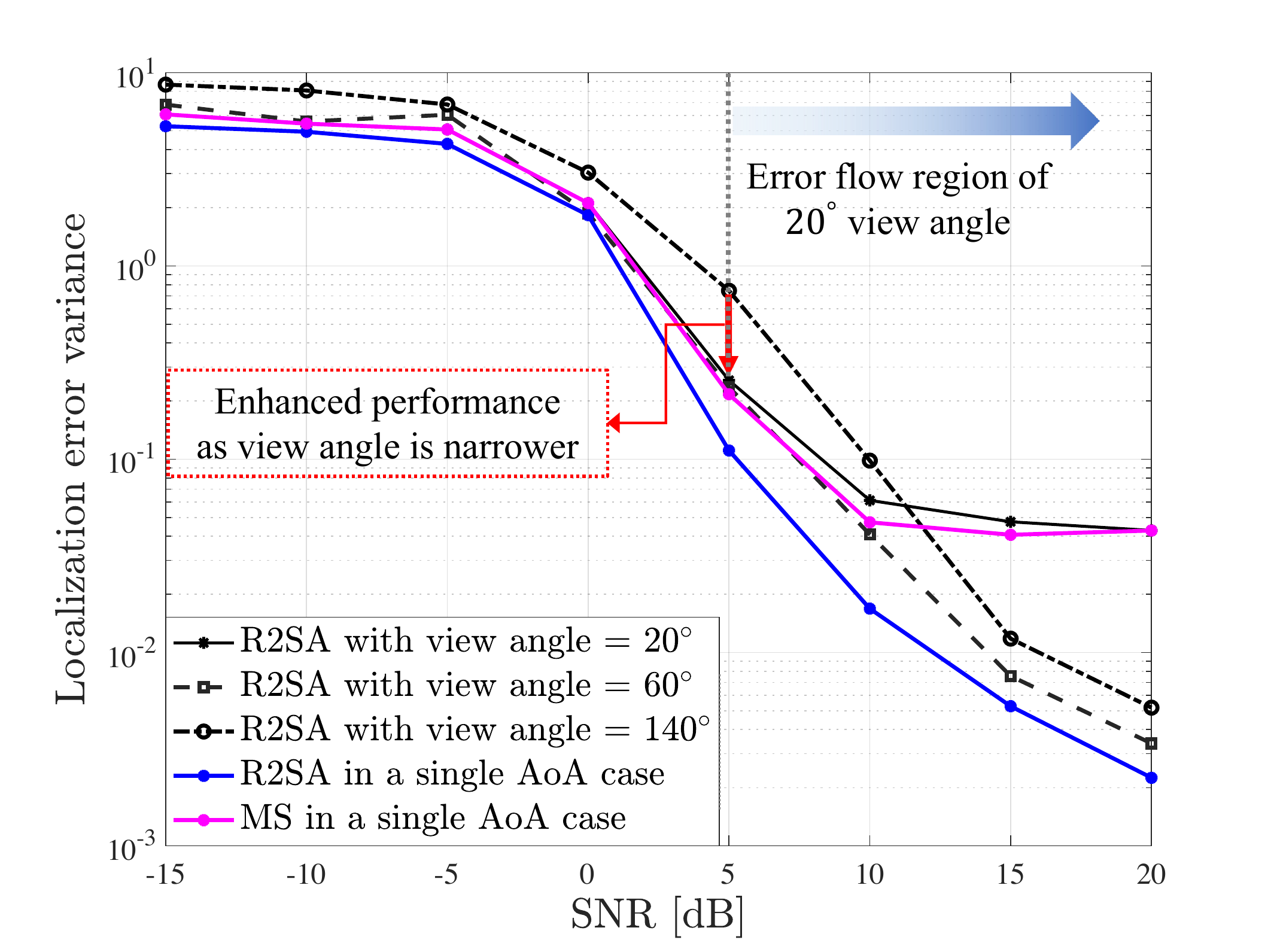}}

\centering
\subfloat[More antenna ($N=61$)]{\includegraphics[width=0.97\linewidth]{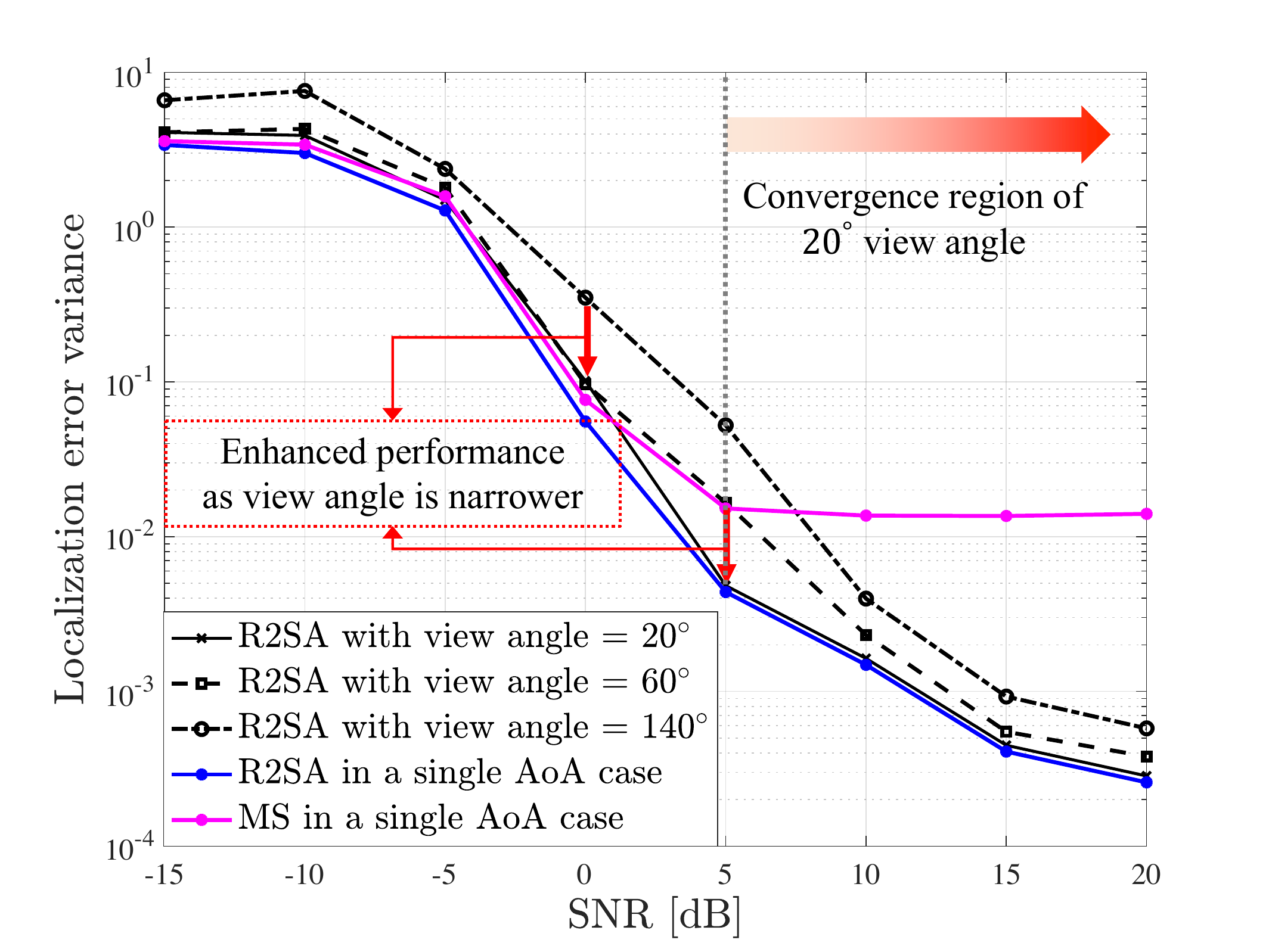}}
\caption{Localization error versus SNR for the angle of view $\in\left\{20,60,140\right\}$ degree a with practical size of the array.}
\label{f11}
\end{figure}

In Fig. \ref{f11}, unlike massive MIMO, we will consider more practical lens-MIMO of 31 and 61 antennas in the mmWave spectrum of 28GHz, where the size of lens-MIMO (i.e., the form factor of lens-MIMO is aperture $\times$ focal length) is $15\times7.5$~cm, and $30\times15$~cm, respectively. We assume that the lens-MIMO of these form factors can be equipped at the front and rear bumpers of vehicles, where their length and width are 5~m and 2~m, respectively as depicted in \cite{vh}, because there will be more chances of collision accidents at the front and back. Further, to exploit a fundamental capability of lens-MIMO that has better AoA estimation performance at the bore sight, we limit the received angular view for $20^{\circ}$, $40^{\circ}$, and $60^{\circ}$. Fig. \ref{f11} presents a comparison of the performances of the R2SA with SIC for different angular views and illustrates the RMSEs of the estimated localization parameters, as defined in \eqref{indl}. The performance is shown for two scenarios: (a) a smaller number of antennas ($N=31$), and (b) a larger number of antennas ($N=61$). The dashed blue and flushed lines are R2SA and MS with a single target in a sub-channel; and the black lines are R2SA for multiple targets in a sub-channel with different view angles, where the total number of vehicles is 8, and the number of target AoAs in a sub-channel is 2.

Fig. \ref{f11}(a) illustrates that the SE-based localization with R2SA achieves the best performance in $20^{\circ}$ view than the others in the SNR range less than 5~dB, but its accuracy shows an error floor as the SNR increases. This is because two received spatial signals are more likely to be focused on a single antenna, making it hard to separate the signals, even with the SIC processing. The R2SA of $60^{\circ}$ view angle shows the best performance in the high SNR region ($\text{SNR}\geq0$). Meanwhile, in Fig. \ref{f11}(b) with larger antenna elements, the R2SA of $20^{\circ}$ view angle with SIC converges to the R2SA of a single target at SNR $=\,5$~dB, where The performance of all three angular views is within a 3 dB difference for the localization error variance of $10^{-3}$. As shown in the results in Section \ref{single}, where the R2SA for a single target satisfied the target requirements (i.e., $0.2\,\text{m}\times1.96\sigma$ for position and $2^{\circ}\times1.96\sigma$ for orientation), it is observed that the R2SA with SIC of the larger number of antennas ($N=61$) guarantees localization accuracy in V2V street intersection scenarios for any angle of views of lens-MIMO in the SNR region of larger than 10~dB. It is noted that the SIC-combined R2SA with a narrower view angle assures the requirements of positioning services.

In the multiple target vehicles case, Figs. 9-11 remark that the R2SA with SIC, using a larger number of antennas, can offer precise V2V positioning services for two target vehicles in the street intersection. This could be achieved with significantly lower complexity compared to conventional methods in ULA systems, as discussed in Section \ref{per}.
\section{Conclusion}
This paper first presented the theoretical limit of the localization of a lens-MIMO for a cooperative vehicle-to-vehicle (V2V) communication in the street intersection, and its results are compared with the conventional uniform linear array (ULA). In the lens-MIMO, we further investigated the characteristic of the received signal to estimate an angle of arrival (AoA) with low complexity. As a result, we proposed a R2SA AoA estimation scheme exploiting the ratio of the two strongest received signals at the antenna elements. For the given AoA estimates, we presented the feasibility of a localization algorithm that solves the sensing equations with respect to the AoA and geometric model. Furthermore, we confirmed that the bounds of position and orientation are affected by the CRLB of AoA, and showed that the localization performance of lens-MIMO is better than the conventional ULA under the specific condition in terms of lens design parameters, such as focal length and lens aperture. 

The simulation results verified that the performance of the localization using R2SA approaches the derived upper bound of the lens-MIMO for a single target vehicle. Furthermore, R2SA outperforms conventional methods such as MUSIC and ML in ULA systems, even though it has lower complexity. It was also confirmed that the proposed localization method, which solves the sensing equations, satisfies the target requirements for the 5G positioning service. In the street intersection scenario with multiple target vehicles, we found that a lens-MIMO with a larger size and narrower angle of view can effectively suppress multi-path interference. Hence, the localization and AoA estimation methods proposed for the street intersection demonstrate promising potential as a framework suitable for 5G/B5G localization use cases that require higher accuracy and lower complexity. As future work, it is promising to extend the proposed method to accommodate multi-path and non-line-of-sight channel parameters in practical environments.

\section{APPENDIX}
\subsection{Appendix A}
\renewcommand{\theequation}{A.\arabic{equation}}
\setcounter{equation}{0}
The first derivative of the $sinc$ function can be represented as
\begin{equation}
    \frac{\delta}{\delta x}\left(\frac{\sin{\pi x}}{\pi x}\right)=\begin{cases}
    0, & \text{for}\, x=0\\
    \frac{\pi x \cos{(\pi x)}-\sin{(\pi x)}}{\pi x^2}, & \text{otherwise.}
    \end{cases}\label{b1}
\end{equation}

Suppose $x=Z$, where $Z\in\mathbb{Z}$ is an arbitrary integer. Then, $\sin{\pi Z}=0$ and 
\begin{equation}
    \frac{\delta}{\delta x}\left(\frac{\sin{\pi x}}{\pi x}\right)\vert_{x=Z}=\begin{cases}
    0, & \text{for}\, Z=0\\
    \frac{1}{Z}, & \text{$Z$ is odd}\\
    -\frac{1}{Z}, & \text{$Z$ is even.}
    \end{cases}\label{b2}
\end{equation}

Then, the upper and lower bounds of $\boldsymbol{\mu}^{\text{T}}_{2}\boldsymbol{\mu}_2$ are simply derived as follows:
\begin{equation}
\sum_{\ell=1}^{\frac{N-1}{2}} \frac{1}{\ell^2}<1+\sum_{\ell=1}^{\frac{N-1}{2}}\frac{1}{\ell(\ell-1)}=2-\frac{2}{N-1}<2,\label{b3}
\end{equation}

\begin{equation}
    \mathbf{\mu}_{2}^\text{T}\mathbf{\mu}_{2}\geq\sum_{\ell=1}^{N}\frac{1}{\ell^2}\geq1.\label{b4}
\end{equation}

By substituting the above inequalities into \eqref{eq23}, the upper and lower bounds of the lens-MIMO's CRLB can be determined as
\begin{equation}
    \frac{f\lambda^2\sigma^2}{4 L^4 \cos^2{\theta}}<\text{CRLB}_{\text{Lens}}(\theta)<\frac{f\lambda^2\sigma^2}{2 L^4 \cos^2{\theta}}.\label{b5}
\end{equation}

For a fair comparison with the CRLB of ULA, let the number of antennas $N$ in \eqref{eq3} be $\left(2\frac{L}{\lambda}+1\right)$. Then, the CRLB of ULA can be reformulated as a function of the lens aperture, it is given by
\begin{equation}
    \text{CRLB}_{\text{ULA}}(\theta)=\frac{3L^2\sigma^2}{L(2L+\lambda)(L+\lambda)d_{\text{ULA}}\cos^2\theta}.\label{b6}
\end{equation}

Using \eqref{b5} and \eqref{b6}, we can readily compare the bounds with and without a lens.
Suppose that $\text{CRLB}_{\text{Lens}} \leq \text{CRLB}_{\text{ULA}}$; this inequality may be represented as a condition in terms of the focal length of the lens. By considering the upper bound of the lens's CRLB in \eqref{b5}, it is given by
\begin{equation}
    f\leq\frac{12L^3}{(2L+1)(L+1)\lambda^4}.\label{b7}
\end{equation}

This completes the proof of \eqref{eq24}.

\subsection{Appendix B}
\renewcommand{\theequation}{B.\arabic{equation}}
\setcounter{equation}{0}
This section focuses on deriving the Fisher information matrix (FIM) elements. The sub-block matrix $[F_{\mathbf{xx}}]_{i,j}$ of the FIM is written as 
\begin{align}
[F_\mathbf{xx}]_{i,j} = \mathbb{E}
\left[\frac{\partial\ln{\mathbf{f}(\boldsymbol{\theta}|\boldsymbol{\eta})}}{\partial x_i}\frac{\partial\ln{\mathbf{f}(\boldsymbol{\theta}|\boldsymbol{\eta})}}{\partial{x_j}} \right].\label{a1}
\end{align}

Note that $\theta_{k,j}$ is independent of $\theta_{k,i}$ when $i \neq j$ by the mmWave assumption, then the entries become zero for $i\neq j$. The two times differentiation of $\mathbf{f}(\boldsymbol{\theta}|\boldsymbol{\eta})$ with respect to $x_i$ and $x_j$ is an exponential function, then we have
\begin{equation}
    [F_{\mathbf{xx}}]_{i,j}=\begin{cases}
    \sum_{j\in\mathcal{V}}\frac{\partial^{2}\ln\mathbf{f}(\theta_{k,j}|\mathbf{p},\omega)}{\partial^{2}x^{2}_k}, & \text{for}\,\, i=j,\\
    0, & \text{otherwise}.\label{AA}
    \end{cases}
\end{equation}

Considering the second derivative, we can derive the diagonal entries of FIM as follows:
\begin{align}
&\frac{\partial^{2}\ln{\mathbf{f}(\theta_{k,j}|\mathbf{p}_{k},\mathbf{p}_{j},\omega_{k,j})}}{\partial{x_k^2}}\\
&=\frac{1}{\sqrt{2\pi\sigma_{n_{k,j}}^{2}}}\frac{\partial}{\partial{x_k}}\left[\frac{\partial}{\partial{x_k}}\left(- \frac{(\theta_{k,j}-\alpha_{k,j})^{2}}{2\sigma_{n_{k,j}}^{2}} \right) \right]\label{a3}\\
&=\frac{1}{\sqrt{2\pi\sigma_{n_{k,j}}^{2}}}\frac{\partial}{\partial{x_k}}\left[\left(\frac{(\theta_{k,j}-\alpha_{k,j})} {\sigma_{n_{k,j}}^{2}} \right) \frac{\partial{\alpha_{k,j}}}{\partial{x_k}} \right]\label{a4}\\
&=\frac{1}{\sqrt{2\pi\sigma_{n_{k,j}}^{2}}}\frac{\partial}{\partial{x_k}}\left[\left(\frac{(\theta_{k,j}-\alpha_{k,j})} {\sigma_{n_{k,j}}^{2}} \right)  \frac{(y_j-y_k)}{d_{k,j}^{2}} \right]\label{a5}\\
&= \mathbf{A}\left[-\frac{(y_j-y_k)^{2}}{d_{k,j}^{4}}+(\theta_{k,j}-\alpha_{k,j})\frac{\partial}{\partial{x_k}}{\frac{(y_j-y_k)}{d_{j,k}^{2}}}\right],\label{a6}
\end{align}
where $\mathbf{A}=\frac{1}{\sigma_{n_{k,j}}^{2}\sqrt{2\pi\sigma_{n_{k,j}}^{2}}}$ is the constant term. In \eqref{a4}, the derivative of the geometric form $\alpha_{k,j}$ is given by
\begin{equation}
    \begin{split}
       \frac{\partial\alpha_{k,j}}{\partial x_k}&=\frac{\partial}{\partial x_k}\left(\tan^{-1}{\left(\frac{y_j - y_k}{x_j - x_k}\right)}-\omega_{k,j}\right)\\
       &=\frac{y_j-y_k}{(y_j-y_k)^2 + (x_j-x_k)^2}=\frac{y_j-y_k}{d_{k,j}^{2}}.\label{a7}
    \end{split}
\end{equation}

Since $\mathbb{E}[\theta_{k,j}]=\alpha_{k,j}$, we have the $k$-th diagonal entry by substituting \eqref{a6} into \eqref{a1} as follows:
\begin{equation}
F_\mathbf{xx}(k,k)=\sum_{j\in\mathcal{V}}\frac{1}{\sqrt{2\pi\sigma_{n_{k,j}}^{2}}}\frac{1}{\sigma_{n_{k,j}}^{2}}\frac{(y_j-y_k)^{2}}{d_{k,j}^{4}}.\label{a8}
\end{equation}

This completes the proof of \eqref{eq11}.
\bibliographystyle{IEEEtran}
\bibliography{bib.bib}

\end{document}